\documentclass[sn-nature]{sn-jnl}

\usepackage{manyfoot}
\usepackage[version=3]{mhchem}
\usepackage{color}
\usepackage{amsmath,amsfonts,amsthm}
\usepackage{graphicx}

\newcommand{\vett}[1]{{\bf{#1}}} 

\newcommand*\crt[1]{\hat{a}^\dagger_{#1}}
\newcommand*\dst[1]{\hat{a}^{\phantom{\dagger}}_{#1}}

\newcommand*\bzmesh[3]{$#1 \times #2 \times #3$}

\begin{document}

\title{Quantum Computation of Reactions on Surfaces Using Local Embedding}

\author*[1]{\fnm{Tanvi P.} \sur{Gujarati}}\email{tgujarati@ibm.com}

\author*[1]{\fnm{Mario} \sur{Motta}} \email{Mario.Motta@ibm.com}

\author[2]{\fnm{Triet} \sur{Nguyen Friedhoff}}

\author[1]{\fnm{Julia E.} \sur{Rice}}

\author*[3]{\fnm{Nam} \sur{Nguyen}}\email{nam.h.nguyen5@boeing.com}

\author[4]{\fnm{Panagiotis Kl.} \sur{Barkoutsos}}

\author[5]{\fnm{Richard J.} \sur{Thompson}}

\author[6]{\fnm{Tyler} \sur{Smith}}

\author[7]{\fnm{Marna} \sur{Kagele}}

\author[8]{\fnm{Mark} \sur{Brei}}

\author[1]{\fnm{Barbara A.} \sur{Jones}}

\author[5]{\fnm{Kristen} \sur{Williams}}

\affil[1]{\orgdiv{IBM Quantum}, \orgname{IBM Research - Almaden}, \orgaddress{\city{San Jose}, \postcode{95120}, \state{California}, \country{USA}}}

\affil[2]{\orgdiv{IBM Quantum}, \orgname{T. J. Watson Research Center}, \orgaddress{\city{Yorktown Heights}, \postcode{10598}, \state{New York}, \country{USA}}}

\affil[3]{\orgdiv{Integrated Vehicle Systems, Applied Mathematics}, \orgname{Boeing Research \& Technology}, \orgaddress{\street{Street}, \city{Huntington Beach}, \postcode{92647}, \state{California}, \country{USA}}}

\affil[4]{\orgdiv{IBM Quantum}, \orgname{IBM Research – Zurich}, \orgaddress{\postcode{8803}, \city{R\"{u}schlikon}, \country{Switzerland}}}

\affil[5]{\orgdiv{Integrated Vehicle Systems, Applied Mathematics}, \orgname{Boeing Research \& Technology}, \orgaddress{\street{Street}, \city{Huntsville}, \postcode{35824}, \state{Alabama}, \country{USA}}}

\affil[6]{\orgdiv{Integrated Vehicle Systems, Applied Mathematics}, \orgname{Boeing Research \& Technology}, \orgaddress{\street{Street}, \city{Orlando}, \postcode{32826}, \state{Florida}, \country{USA}}}

\affil[7]{\orgdiv{Tech Vis and Integration, Global Technology}, \orgname{Boeing Research \& Technology}, \orgaddress{\street{Street}, \city{Tukwila}, \postcode{98108}, \state{Washington}, \country{USA}}}

\affil[8]{\orgdiv{BSC Analytics}, \orgname{Chemical Technology}, \orgaddress{\city{North Charleston}, \postcode{29456}, \state{South Carolina}, \country{USA}}}

\abstract{Modeling electronic systems is an important application for quantum computers.
In the context of materials science, an important open problem is the computational description of chemical reactions on surfaces.
In this work, we outline a workflow to model the adsorption and reaction of molecules on surfaces using quantum computing algorithms.
We develop and compare two local embedding methods for the systematic determination of active spaces. These methods are automated and based on the physics of molecule-surface interactions and yield systematically improvable active spaces.
Furthermore, to reduce the quantum resources required for the simulation of the selected active spaces using quantum algorithms, we introduce a technique for exact and automated circuit simplification. This technique is applicable to a broad class of quantum circuits and critical to enable demonstration on near-term quantum devices.
We apply the proposed combination of active-space selection and circuit simplification to the dissociation of water on a magnesium surface using classical simulators and quantum hardware. Our study identifies reactions of molecules on surfaces, in conjunction with the proposed algorithmic workflow, as a promising research direction in the field of quantum computing applied to materials science.}

\maketitle

\section{Introduction}

The accurate computational description of correlated electrons in materials is an outstanding research challenge. Quantitative simulations of electronic wavefunctions are essential for accurate and predictive calculations
of properties, such as the rates at which industrially relevant or biologically and environmentally hazardous reactions occur. However, it requires a sufficiently accurate solution of an underlying Schr\"{o}dinger equation.
The combinatorial growth of the many-electron Hilbert space, along with the high
degree of entanglement produced by electron-electron interaction and Fermi statistics means that the computational cost of exactly solving the Schr\"{o}dinger equation scales combinatorially with system size, a formidable obstacle that has led to the development of approximate numerical techniques.
Methods based on density functional theory (DFT) have had an enormous impact 
on materials science, but become sensitive to the underlying approximations 
in presence of static electron correlation~\cite{cohen2008insights,mori2008localization,cohen2008fractional,jones1989density}.
Therefore, a topic of considerable interest is the development of systematic numerical approaches that are chemically realistic and fundamentally many-body.

These approaches include algorithms for quantum computers, which have the potential
to accurately and efficiently simulate correlated electronic systems from first principles~\cite{lloyd1996universal,georgescu2014quantum,cao2019quantum,bauer2020quantum}.
Quantum algorithms, based on quantum resource estimates and coupled with classical simulations~\cite{Wecker2014,Reiher2016PNAS,Elfving2020,Goings2022}, are projected to deliver results that are competitive with classical methods in both accuracy and cost for specific classes of correlated electronic problems.
Common to these problems is the presence of static correlation from electrons and orbitals in a spatially local region, and dynamical correlation from the remaining degrees of freedom.

Many important applications in the electronics, aerospace,  automobile, and defense sectors feature a spatially localized region in which electron correlation effects are expected to be more important than in the rest of the system. An example is the corrosion on metallic surfaces, which is initiated by the adsorption of reactants (atoms or molecules from the environment) on a spatially local portion of the surface. 
In such a situation, it is chemically justified to treat only a portion of the system with an accurate many-body method, and the rest of the system with a less expensive mean-field method. This feature makes reactions on surfaces an especially compelling target for studies on near-term quantum devices, in conjunction with techniques to select relevant degrees of freedom and reduce the budget of quantum simulations.

Here, we propose an algorithmic workflow to simulate reactions on surfaces on quantum computers. The proposed workflow comprises an embedding method specifically designed for reactions of molecules on surfaces, and a circuit simplification technique to facilitate experiments on near-term quantum devices.

First, we developed and compared two methods that are used to rank and select active-space orbitals based on (i) their contribution to the difference between the DFT electronic density of the system and the superimposed DFT electronic densities of the constituent surface and adsorbate and (ii) their effect on the ground-state active-space energy.
Second, we solved the Schr\"{o}dinger equation in the active space using the variational quantum eigensolver~\cite{peruzzo2014variational}. To achieve this goal, it was necessary to evaluate the expectation value of the active-space Hamiltonian over a quantum circuit. We simplified and economized this operation by employing the algebraic properties of Clifford transformations. This allowed for the construction of an equivalent circuit with fewer qubits and gates, and lower depth compared to the original one.

We illustrated the proposed workflow on a step in the corrosion reaction of magnesium by water~\cite{williams2015first, williams2016modeling, wurger2020first, yuwono2019aqueous, Yuwono2016, Limer2017}. We discussed the underlying approximations and assessed their impact on the accuracy of the computed properties. Finally, we demonstrated the proposed workflow using IBM's quantum hardware.

\section{Results}

\subsection{Chemical Reaction}

The corrosion of magnesium in water or aqueous environment proceeds by an electrochemical reaction that produces magnesium hydroxide and hydrogen gas. While the overall corrosion reaction is well-known,
\begin{equation}
\ce{Mg + 2H2O <=>> \text{\ce{Mg(OH)2}} + H2}
\;,
\label{eq:Mg_cor_main}
\end{equation}
the detailed mechanisms of hydrogen evolution reactions on a magnesium surface are a topic of ongoing investigation~\cite{williams2015first,williams2016modeling,yuwono2019aqueous,wurger2020first, esmaily2017fundamentals}.

\begin{figure*}[t!]
	\centering
	\includegraphics[trim={0 0 0 1.5cm},clip,width=\textwidth]{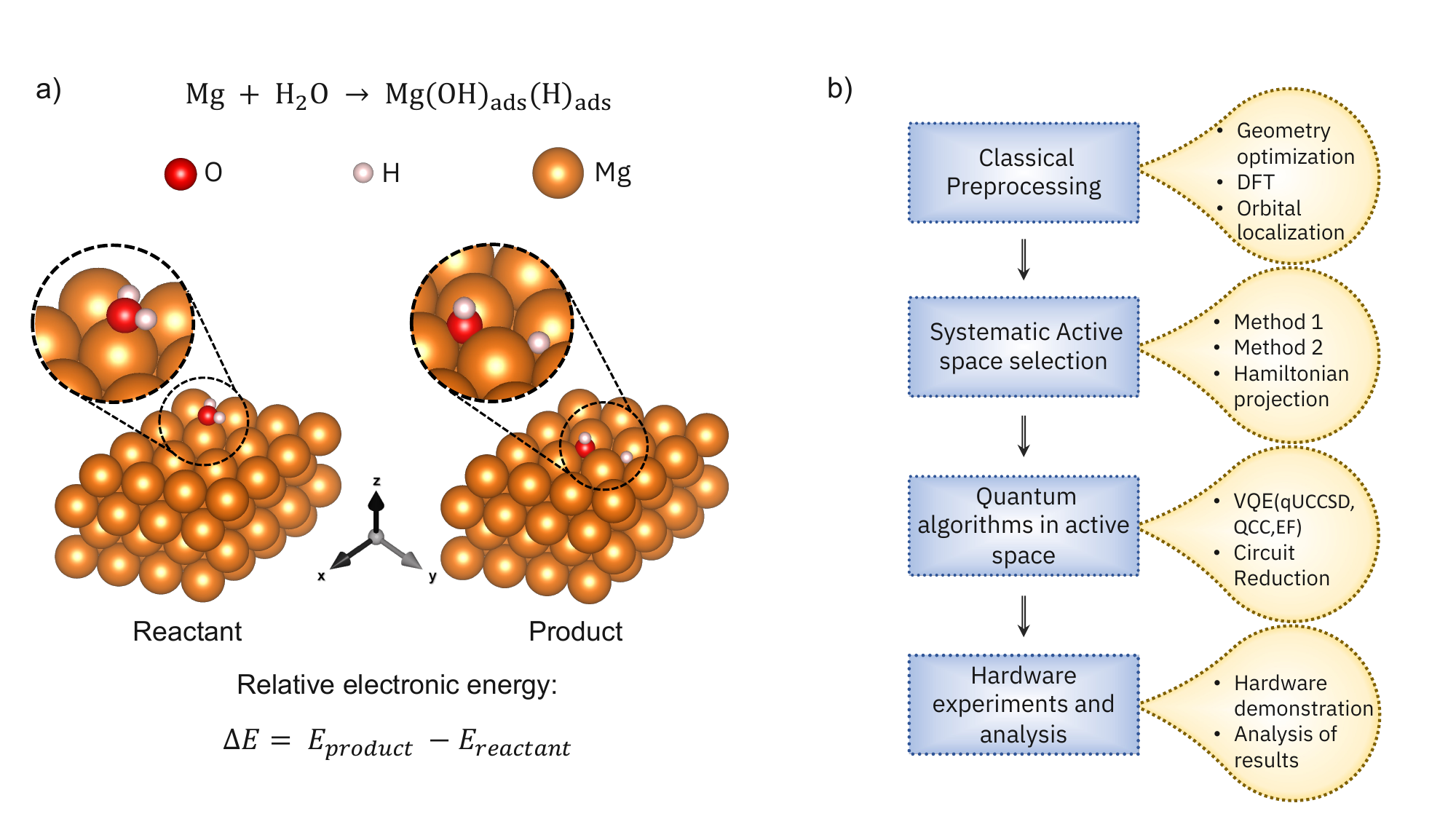}
	\caption{\textbf{Description of the chemical reaction and the workflow.} (a) reaction for splitting of water on a magnesium surface, including schematics of the optimized structures for the reactant and product. (b) Summary of the different steps involved in the workflow. Each step of the workflow is described in detail in the text.}
	\label{fig:workflow}
\end{figure*}

Williams et al~\cite{williams2016modeling} proposed a detailed reaction scheme connecting the steps of initial water dissociation on Mg surface with the final step of $\ce{H_{2}}$ evolution via a Tafel mechanism~\cite{Sharifi2013, williams2016modeling} in the presence of adsorbed OH$_\text{ads}$ and H$_\text{ads}$ species using modeling based on DFT. The suggested reaction mechanism was shown to be a concerted reaction involving multiple water molecules. The first reaction studied in the process was the splitting of a single \ce{H_{2}O} molecule creating adsorbed H$_\text{ads}$ and OH$_\text{ads}$ moiety,
\begin{equation}
\ce{Mg + H_{2}O -> Mg(OH_{ads})(H_{ads}) }
\;.
\label{eq:Mg_singleH2O}
\end{equation}
While many steps are involved in the study of the hydrogen evolution process, as discussed in the Supplementary Information (SI), in this work we focused on modeling the chemical reaction in Eq. (\ref{eq:Mg_singleH2O}) using the workflow described in Fig.~\ref{fig:workflow}. 
In particular, we computed the electronic energy difference between the reactant and product,
\begin{equation}
\Delta E = E_{\mathrm{product}} - E_{\mathrm{reactant}}
\;
\label{eq:DeltaE}
\end{equation}
Eq.~\eqref{eq:DeltaE} is an important quantity since it is used in the determination of thermodynamic quantities such as the enthalpy or the Gibbs free energy of reaction. In addition to thermodynamics, it is important to characterize the kinetics of surface reaction processes. Determining the kinetics of Eq. (\ref{eq:Mg_singleH2O}) involves calculating the activation energy (i.e., the difference between the transition state and reactant energy). Williams et al~\cite{williams2016modeling} found that the activation energy for \ce{H_2O} dissociation on Mg is 1.31 eV for a single \ce{H_2O} molecule and 1.06 eV for a concerted reaction involving multiple \ce{H_2O} molecules. Although we did not calculate activation energies in this study, we plan to explore transition states in future research.

\subsection{Classical Pre-processing}

The workflow in Fig.~\ref{fig:workflow}b starts with classical pre-processing. We obtained optimized geometries of the reactants and products using DFT with periodic boundary conditions (PBC). Schematics of these structures are shown in Fig.~\ref{fig:workflow}a. In the optimized structure for the reactant, the water molecule is adsorbed on the surface with the oxygen atom situated 2.4~\r{A} above an atop site. In the optimized structure for the product, the water molecule is split such that OH$\rm{_{ads}}$ and H$\rm{_{ads}}$ are co-adsorbed at nearest-neighbor \textit{fcc} sites. 

We carried out the simplest PBC calculations at the center of the Brillouin point ($\vett{\Gamma}$ point), where the Hamiltonian is time-reversal-symmetric. However, $\vett{\Gamma}$ point calculations are known to converge slowly and non-monotonically to the thermodynamic limit of infinite system size at zero temperature. To achieve better convergence, in this work, we used twist-averaged boundary conditions (TABC)~\cite{lin2001twist} as an economical alternative to 
full Brillouin zone sampling~\cite{mcclain2017gaussian,zhang2018auxiliary,motta2019hamiltonian}. Within TABC, the expectation value of an operator $B$ is averaged over a mesh of $N_k$ points $\vett{k}_i$ in the Brillouin zone,
$\langle B \rangle = \frac{1}{N_k} \sum_{ i=1}^{N_k} B(\vett{k}_i)$.

At the optimized geometries, we computed the energy difference in Eq.~\eqref{eq:DeltaE} at the DFT level of theory 
(see Methods).
The DFT calculations yielded $\Delta E = -1.91$ eV at the $\vett{\Gamma}$ point. By comparison, Williams et al~\cite{williams2016modeling} reported $\Delta E = -1.78$ eV. This difference originates from the different optimized geometries, basis sets, and DFT functionals used in the two studies. As a verification, we quantified the basis set superposition error affecting our DFT calculations using the counterpoise correction~\cite{BSSE1999}, which yielded $\Delta E_{\mathrm{cp}} = -1.77$ eV, in better agreement with the value obtained by Williams et al in a large plane-wave basis.

To quantify the finite-size error on DFT energies, we performed TABC calculations with \bzmesh{2}{2}{1} and \bzmesh{4}{4}{1} Monkhorst-Pack~\cite{monkhorst1976special} meshes of $\vett{k}$ points, which increased $\Delta E$ by 0.093 and 0.262 eV compared against the $\vett{\Gamma}$ point.

\subsubsection{Active-Space Selection}

For certain chemical problems, reasonably accurate results can be achieved by correlating a limited number of electrons and orbitals through an active-space calculation~\cite{werner1985second,knowles1985efficient}. In general, an active space of valence electrons and orbitals is most desirable, and further reductions are acceptable when justified by chemical grounds. In particular, all orbitals responsible for static correlation have to be included in the active space~\cite{keller2015selection}.

Previous work showed that, for some systems comprising small molecular adsorbates on surfaces, electronic correlation is primarily associated with a limited number of orbitals and electrons~\cite{Baumann2015Fe, Rau2014Co, Cyrus2007Co, Oliver2015Ni}. These observations suggest the possibility of constructing compact active spaces for reactions on surfaces. Such a construction should be automated~\cite{sayfutyarova2017automated,stein2016automated} and physics-based. Furthermore, active spaces should be systematically improvable, to allow convergence of computed properties. 

In this work, we designed and compared two active-space construction strategies satisfying the above requirements. The starting point of both methods was the separate localization of occupied and virtual DFT orbitals (see Methods), and their projection onto an active region~\cite{eskridge2019local} comprising the molecules participating in the reaction and a small portion of the surface.

\begin{figure*}[t!]
	\centering
	\includegraphics[width=\textwidth]{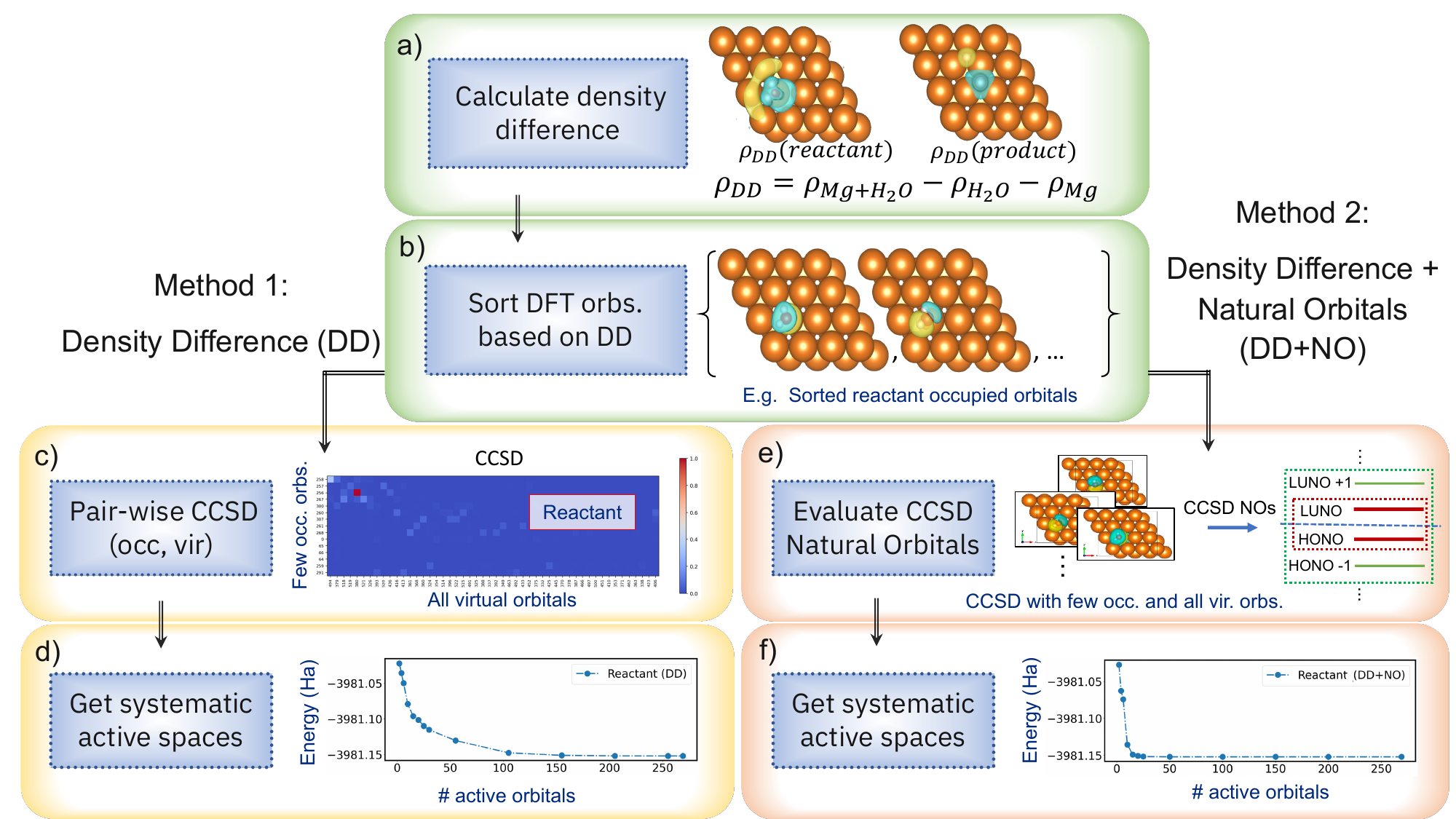}
	\caption{\textbf{Active-space selection methods based on the electron density difference.} The top two blocks (a,b) are common to both methods. The two blocks on the left (c,d) illustrate the steps of the first method, denoted Density Difference (DD). The two blocks on the right (e,f) describe the steps of the second method, denoted Density Difference + Natural Orbitals (DD+NO).}
	\label{fig:act_space_methods}
\end{figure*}

\paragraph{Method 1 - Based on Density Difference (DD):} This method ranks occupied DFT orbitals according to their contribution to the difference
\begin{equation}
\rho_{\mathrm{DD}}(\vett{x})  = \rho_{\mathrm{Mg+H_{2}O}}(\vett{x}) - \rho_{\mathrm{H_{2}O}}(\vett{x}) - \rho_{\mathrm{Mg}}(\vett{x})
\label{eq:chrg_den_diff}
\end{equation}
between the DFT electronic density of the full system and the sum of the DFT electronic densities of adsorbate and slab. More specifically,
we multiplied $\sqrt{\rho_{\mathrm{DD}}(\vett{x})}$ times the absolute value of each localized occupied DFT orbital $| \psi_i(\vett{x}) |$, integrated this product over space, and retained the five (as many as the valence occupied orbitals of $\ce{H_{2}O}$) orbitals with the highest integrated overlaps (see also Fig.~\ref{fig:act_space_methods}ab and Methods).
On the one hand, this method provides a simple and inexpensive way of ranking occupied DFT orbitals. On the other hand, the ranking of virtual DFT orbitals is more subtle, because they do not significantly contribute to Eq.~\eqref{eq:chrg_den_diff}. For each retained occupied and virtual DFT orbital, respectively $\psi_i$ and $\psi_a$, we computed the CCSD (coupled-cluster singles and doubles) energy in a (2e,2o) active space spanned by $\psi_i$ and $\psi_a$ (see Fig.~\ref{fig:act_space_methods}c). It is worth noting that, for two-electron systems, CCSD is exact.
We then sorted pairs $(i,a)$ according to the value of the (2e,2o) CCSD energy, and retained the highest-ranking virtual orbitals. This method is efficient in terms of the required classical resources and yields active-space ground-state energies that decrease monotonically with increasing active-space size (see Fig.~\ref{fig:act_space_methods}d).

\paragraph{Method 2 - Based on Density Difference and CCSD Natural Orbitals (DD+NO):} As seen in Fig.~\ref{fig:act_space_methods}d, the energy converges slowly with active-space size. Convergence improves considerably using natural orbitals~\cite{khedkar2019active,abrams2004natural}.
In the DD+NO method, we carried out a CCSD calculation in the active space spanned by the five highest-ranking occupied DFT orbitals (determined as in the DD method) and all virtual orbitals. We then constructed natural orbitals as eigenvectors of the CCSD one-particle density matrix, sorted them in decreasing order of occupation number.
For the problems studied here, occupation numbers were either close to 2 or to 0, so we could unambiguously divide orbitals as high- and low-occupancy (in systems with strong correlation, there is a set of natural orbitals with fractional occupation numbers, and it is necessary to include them in the active space).
We sorted natural orbitals in decreasing order of occupation number and defined the highest-occupied and lowest-unoccupied natural orbitals (HONO and LUNO respectively) as the orbitals with indices $N_e/2$ and $N_e/2+1$, where $N_e$ is the number of electrons. We then constructed active spaces spanning orbitals between HONO$-a$ and LUNO$+b$ with $a=\min(k,N_e/2-1)$ and $b=\min(k,N_o-N_e/2-1)$ where $k \geq0$ is an integer and $N_o$ the total number of orbitals (see Fig.~\ref{fig:act_space_methods}e).

\begin{figure}[h]
	\centering
	\includegraphics[width=\textwidth]{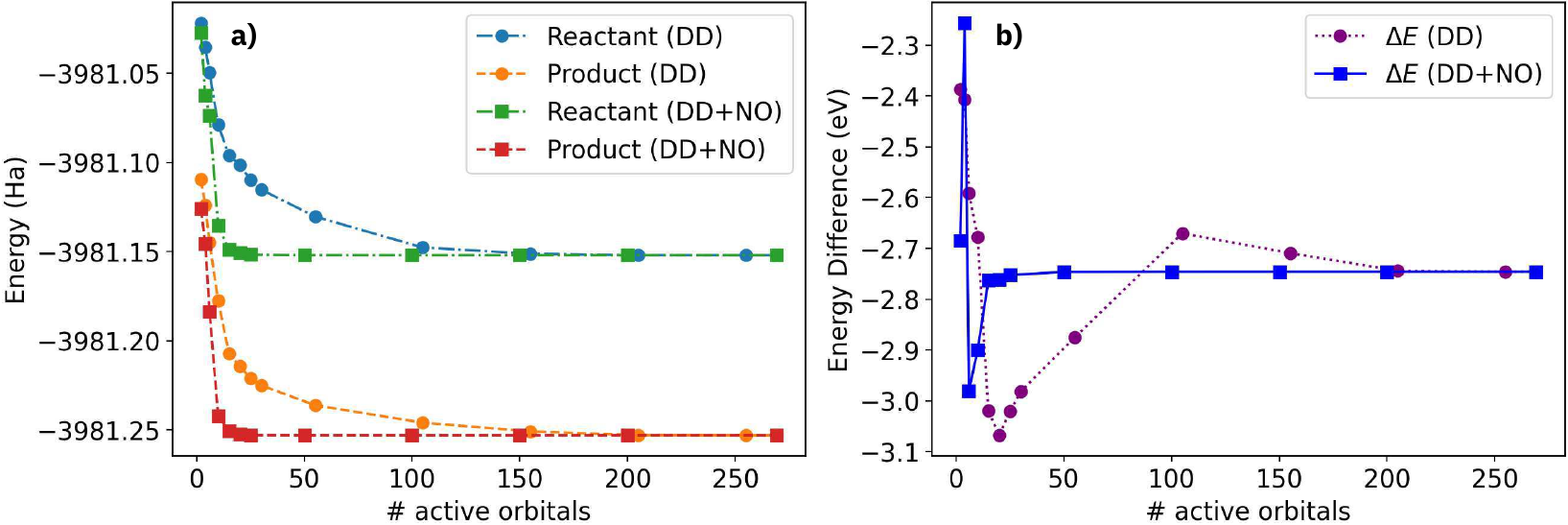}
	\caption{\textbf{Comparison of active-space selection methods.} CCSD total energies (a) and energy differences $\Delta E$ (b) calculated in active spaces constructed with DD and DD+NO methods.}
	\label{fig:comparison_act_spc}
\end{figure}

\paragraph{Comparison:} In Fig.~\ref{fig:comparison_act_spc}, we compared DD and DD+NO methods by computing the CCSD total energies of reactant and product (left) and the corresponding energy difference (right) as a function of active-space size (see Methods).
We considered active spaces of up to 269 orbitals (5 of which are occupied) out of the 588 orbitals in the underlying Gaussian basis set and evaluated energies and energy differences at the $\vett{\Gamma}$ point. DD+NO total energies and energy differences converged faster than their DD counterparts. In particular, only 15-20 natural orbitals are needed to converge $\Delta E$. 
Therefore, in the remainder of this work, we used active spaces constructed with DD+NO. We remark that, while the DD+NO method offers faster ground-state energy convergence, it is considerably more expensive than the DD method, as it requires correlated calculations, which become challenging for bases of hundreds of orbitals. The DD and DD+NO methods are not mutually exclusive but can be used as complementary approaches, for example, the DD method can be used to rank occupied orbitals and identify a subset of relevant virtual DFT orbitals, that can then be treated with DD+NO.

\subsection{Quantum algorithms in the active space}

After identifying the active-space orbitals for each $\vett{k}$ point, we froze the remaining orbitals and projected the Born-Oppenheimer Hamiltonian onto the active space with a standard procedure \cite{eskridge2019local, Gao2021}. 
We represented the active-space Hamiltonian in second quantization, and mapped it to a qubit operator using conventional fermion-to-qubit mappings, namely Jordan-Wigner (JW) and parity with two-qubit reduction (P2QR)~\cite{Jordan1928, ortiz2001quantum, BRAVYI2002210}.

\subsubsection{Variational Quantum Eigensolver (VQE)} 

We performed active-space simulations using the VQE method, wherein the ground-state wavefunction and energy, $E_{0}$, are approximated by variationally optimizing a parameterized wavefunction ansatz $|\psi(\boldsymbol{\theta})\rangle$,
\begin{equation}
E_{\mathrm{VQE}} = \min_{\boldsymbol{\theta}} \langle \psi(\boldsymbol{\theta})|H| \psi(\boldsymbol{\theta})\rangle
\;.
\end{equation}
The function $E_{\mathrm{VQE}}$ is evaluated on a quantum computer, and the parameters $\boldsymbol{\theta}$ are optimized on a classical computer.
The quality of a VQE calculation, and particularly the difference $E_{\mathrm{VQE}}-E_{0}$, depends on the VQE ansatz and the convergence of the optimization procedure. Literature~\cite{Rice2021, Barkoutsos2018, Grimsley2019, Sokolov2020ooUCCSD, Ollitrault2020qEOM, Gao2021} indicates that VQE applied to small systems can yield energies close to those of exact diagonalization in the active space, known as complete active space configuration interaction (CASCI).

Here, we studied the performance of the VQE algorithm using a Trotterized implementation of Unitary CCSD (qUCCSD)~\cite{Barkoutsos2018}, Entanglement Forging (EF)~\cite{eddins2022doubling}, and Qubit Coupled Cluster (QCC)~\cite{QCC2018}. See Methods for more information.

\begin{figure}[h]
	\centering
	\includegraphics[width=\textwidth]{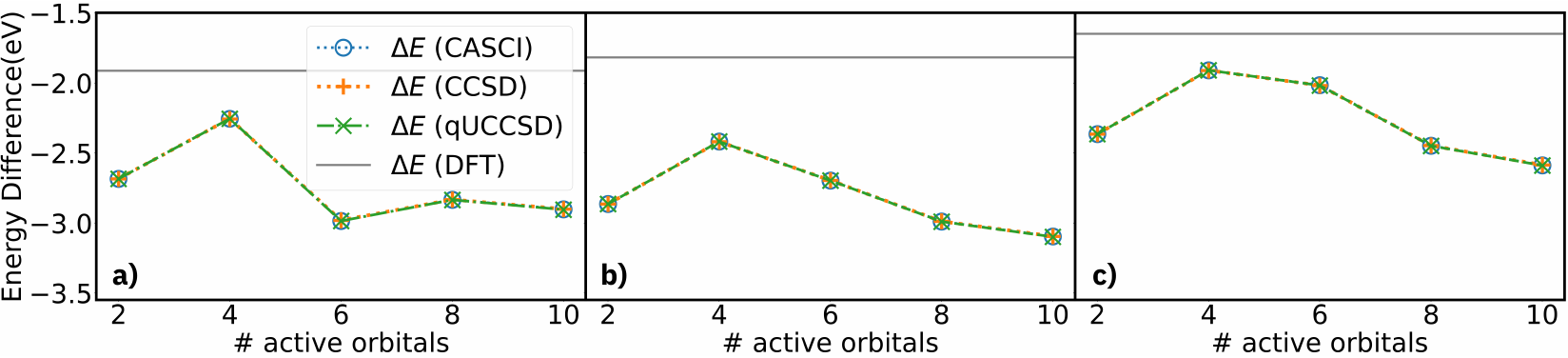}
	\caption{\textbf{Active-space energy differences.} Energy differences $\Delta E$ with CASCI, CCSD, and qUCCSD evaluated over active spaces of 2 to 10 CCSD natural orbitals constructed with the DD+NO method. $\Delta E$ is computed at the $\vett{\Gamma}$ point (a) and with TABC over \bzmesh{2}{2}{1} (b) and \bzmesh{4}{4}{1} (c) Monkhorst-Pack grids. DFT energy differences are shown for reference in grey.}
	\label{fig:small_act_sp}
\end{figure}

We begin our analysis in Fig.~\ref{fig:small_act_sp}, where we compare $\Delta E$ from qUCCSD against CCSD and CASCI. We illustrate the impact of going beyond the $\vett{\Gamma}$ point (left) via TABC over Monkhorst-Pack grids of \bzmesh{2}{2}{1} (middle) and \bzmesh{4}{4}{1} (right) $\vett{k}$ points. 
qUCCSD, CCSD, and CASCI are indistinguishable for all active-space sizes and $\vett{k}$ point meshes. DFT energy differences are also reported for comparison.

The qUCCSD ansatz is accurate but expensive, featuring circuits of depth scaling as $O(N^{4})$, where $N$ is the number of active-space orbitals~\cite{Motta2020transcorrelated}. This fact motivated the development of ansatzes that reduce computational cost while retaining accuracy. Here, we employed one such ansatz, QCC~\cite{QCC2018,QCC2020, Tang2021qAdaptVQE}, which applies exponentials of suitably-chosen Pauli operators $P_1 \dots P_m$ (see Methods) to the Hartree-Fock state,

\begin{equation}
|\psi_{\mathrm{QCC}}(\boldsymbol{\theta})\rangle 
= 
e^{-i\theta_{m}P_m} \dots e^{-i\theta_{1}P_{1}}
|\psi_{\mathrm{HF}}\rangle 
\label{eq:QCC1}
\;.
\end{equation}

\begin{figure}[h]
	\centering
	\includegraphics[width=0.8\textwidth]{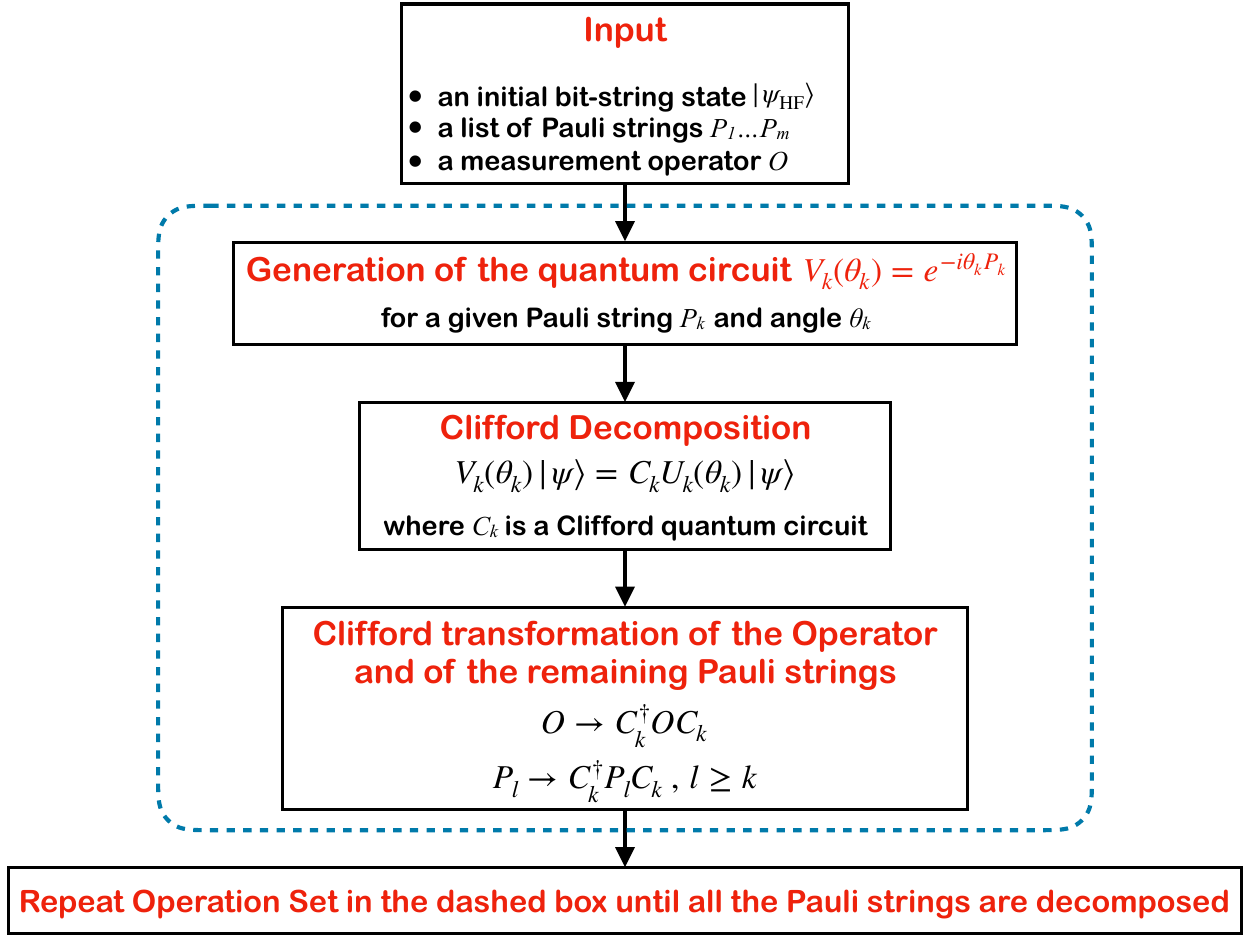}
	\caption{{\bf{Circuit reduction.}} Flowchart of the circuit reduction technique for circuits comprising initialization of qubits in a computational basis state $\psi_{\mathrm{HF}}$ (or bitstring), a product of exponentials of Pauli operators $P_k$ (or Pauli strings), and the measurement of a Pauli operator $O$.}
	\label{fig:circ_red_wf}
\end{figure}

\subsubsection{Circuit reduction}
\label{sec:circuit_reduction}

Evaluating the QCC energy requires computing expectation values of Pauli operators $O$ over a state of the form Eq.~\eqref{eq:QCC1},
\begin{eqnarray}
\langle \psi(\boldsymbol{\theta})|O|\psi(\boldsymbol{\theta})\rangle = \langle \psi_{\mathrm{HF}}|e^{i\theta_{1}P_{1}}...e^{i\theta_{m}P_m}|O|e^{-i\theta_{m}P_m}...e^{-i\theta_{1}P_{1}}|\psi_{\mathrm{HF}}\rangle \;,
\label{eq:Pauli_exponential}
\end{eqnarray}
which is challenging on near-term quantum devices due to the high number of qubits, gates, high circuit depth, and limited qubit connectivity. Here, we devised a circuit simplification technique that significantly reduced the quantum resources required for computing Eq.~\eqref{eq:Pauli_exponential}.

(i) As a preliminary step, we permuted qubits in the register so that Pauli operators $P_k$ act non-trivially, e.g., on the rightmost qubits, prioritizing $P_k$ over $P_{k+1}$.
(ii) We then represented the state $e^{-i\theta_{1}P_{1}} |\psi_{\mathrm{HF}}\rangle$ as $C_1 U_1(\theta_1) |0\rangle$, where $C_1$ is a Clifford transformation. We constructed $C_1$ using a combination of standard circuit identities, such as commuting a Hadamard gate through a CNOT gate~\cite{Nam2018,Bravyi2021cliffordcircuit,Cowtan2020}, as discussed in the SI.
(iii) We used $C_1$ to perform a similarity transformation on the Pauli operators $P_2, \dots, P_m$, and $O$ without altering the expectation value in Eq.~\eqref{eq:Pauli_exponential},
\begin{eqnarray}
\langle \psi(\boldsymbol{\theta})|O|\psi(\boldsymbol{\theta})\rangle 
&=& 
\langle \psi_{\mathrm{HF}}|e^{i\theta_{1}P_{1}} \dots e^{i\theta_{m}P_m}|O|e^{-i\theta_{m}P_m} \dots e^{-i\theta_{1}P_{1}}|\psi_{\mathrm{HF}}\rangle \\
&=& 
\langle 0|U_{1}^{\dagger}(\theta_{1}) e^{i\theta_{2}P_2^\prime} \dots e^{i\theta_{m}P_m^\prime}|O^\prime|e^{-i\theta_{m}P_m^\prime} \dots e^{-i\theta_{2}P_2^\prime} U_{1}(\theta_{1})|0\rangle
\;,
\label{eq:clifford_transformation}
\end{eqnarray}
where $P^\prime_k = C_1^\dagger P_k C_1$ and $O^\prime = C_1^\dagger O C_1$
are Pauli operators that can be determined efficiently~\cite{gottesman1998heisenberg} on a classical computer. In Eq.~\eqref{eq:clifford_transformation} we removed the Clifford transformation $C_1$ from the quantum circuit executed on hardware by applying similarity transformations to $P_k$ and $O$. Thanks to this removal, and the fact that the circuit $U_1(\theta_1)$ has by construction shorter depth and fewer gates than $V_1(\theta_1)$, the simplified circuit has shorter depth and fewer gates than the original one.
(iv) we repeated the previous two steps for each Pauli operator in the circuit. At the end of step (iv), we determined whether the reduced circuit acts trivially on one or more qubits, and removed such qubits (if any) from the calculation. 
A detailed workflow is shown in Fig.~\ref{fig:circ_red_wf} and a complete example is shown in the SI. 

By applying this technique, the quantum resources to simulate the QCC ansatz
are significantly reduced, as exemplified in Table~\ref{Table: Circuit_reduction_numbers} for reactant and product at the $\boldsymbol{\Gamma}$ point.

\begin{table}[h]
\begin{tabular}{cccccccc}
 \hline\hline
 \#Natural & \#Pauli & \#CNOTs & \#CNOTs & Depth & Depth & \#Qubits & \#Qubits \\ [0.5ex]
 Orbitals & Strings & (before) & (after) & (before) &  (after) & (before) & (after) \\
 \hline
 2 & 3 & 2/2 & 1/1 & 18/18 & 6/6 &2/2 & 2/2\\
 \hline
 4 & 25 & 130/122 & 39/57 & 256/248 & 60/72 & 6/6 & 6/6\\
 \hline
 6 & 25 & 180/210 & 62/86 & 306/336 & 70/81 & 10/10 & 9/9 \\
 \hline
 8 & 25 & 340/346 & 74/47 &  466/472 & 73/61 & 14/14 & 11/8 \\
 \hline
 10 & 25 & 482/396 & 72/72 & 608/522& 55/69 & 18/18 &  13/13\\
 \hline\hline
\end{tabular}
\caption{Number of CNOT gates and circuit depth before and after applying the circuit reduction procedure to QCC ansatzes with varying number of Pauli strings (second column), for reactant/product in active spaces of different sizes (first column).}
\label{Table: Circuit_reduction_numbers}
\end{table}

\subsubsection{Simulations on quantum hardware}

\begin{figure}[h]
	\centering
	\includegraphics[width=\textwidth]{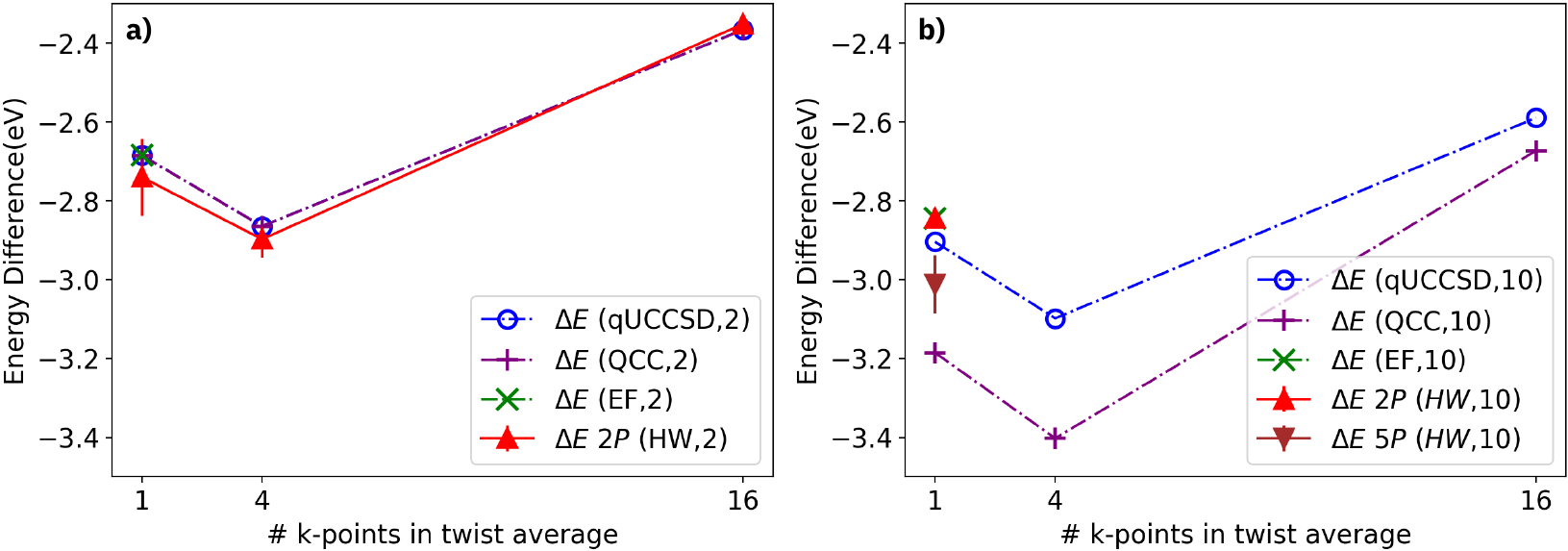}
	\caption{\textbf{Results from quantum algorithms and hardware experiments.} Energy differences $\Delta E$ from noiseless classical simulations and hardware experiments, for active spaces of 2 (a) and 10 (b) natural orbitals from the DD+NO method. For 10-orbital active spaces, we employed QCC with 50 Pauli operators (purple crosses) on classical simulators and QCC with 2 and 5 Pauli operators ($\Delta E~2P$ and $\Delta E~5P$) on quantum hardware.}
	\label{fig:qc_algos}
\end{figure}

In Fig.~\ref{fig:qc_algos} we simulate VQE with QCC and EF ansatzes, using qUCCSD as a reference due to its high accuracy, established in Fig.~\ref{fig:small_act_sp}. 
We start by considering a (2e,2o) active space spanned by the HONO and LUNO orbitals and \bzmesh{n}{n}{1} meshes of $\vett{k}$ points with $n=1,2,4$ (left panel). This active space requires 4 and 2 qubits in JW and P2QR representations, respectively (left panel). Noiseless classical simulations show that various ansatzes are in agreement with each other and with qUCCSD.
Hardware simulations using QCC with 2 Pauli strings as ansatz (red triangles) are statistically compatible with noiseless classical simulations.

We then consider a (10e,10o) active space spanned by the HONO-4 to LUNO+4 orbitals and \bzmesh{n}{n}{1} meshes of $\vett{k}$ points with $n=1,2,4$ (right panel). This active space generally requires 20 and 18 qubits in JW and P2QR representations respectively (left panel).
On noiseless classical simulators, we employed QCC with 50 Pauli strings. On quantum hardware, we implemented QCC with 2 and 5 Pauli strings. We used the circuit reduction technique outlined in the previous section to achieve more economical simulations. In the case of 2 and 5 Pauli strings, reduced circuits acted on 2 and 5 qubits respectively, and had depth 2. The original circuits required, for the reactant/product systems, are (i) 15/14 qubits, 24/20 CNOT gates, and 35/31 circuit depth for QCC with 2 Paulis and (ii) 17/16 qubits, 64/56 CNOT gates and 85/77 circuit depth for QCC with 5 Paulis.
We note that hardware simulations are statistically compatible with noiseless classical simulations using the same quantum circuit.
Furthermore, energy differences computed with QCC depend non-trivially on the number of Pauli operators in the ansatz: in particular, simulations using 50 Pauli operators differ from qUCCSD by roughly $0.1$-$0.3$ eV.
At the $\vett{\Gamma}$ point, we incorporated results from EF for reference. This method can tackle (2e,2o) and (10e,10o) active spaces with 2 and 10 qubits respectively. EF yields results in good agreement with qUCCSD and QCC. However, it should be noted that its current implementation is limited to Hamiltonians with time-reversal symmetry.

Additional information about the performance of the QCC and EF ansatzes is provided in the SI.

\section{Discussion}

Here, we proposed a workflow to simulate reactions of molecules on surfaces with quantum computing algorithms. The proposed workflow comprises an active-space construction and a circuit simplification to economize quantum simulations and facilitate their demonstration on near-term quantum devices.

\paragraph{Embedding/active-space construction:} 
Several methods using quantum algorithms as solvers for one or more active regions have been recently proposed, targeting model systems of strong electronic correlation \cite{Kawashima_2021} and spin defects in semiconductors and insulators \cite{Ma2020, Huang2022, Nan2022}.
An important contribution of our study is the design of automated active-space selection techniques specifically tailored for reactions on surfaces. In this situation, electronic correlation arises primarily from a spatially localized region, suggesting the possibility to construct compact active spaces, but the clear identification of such region is non-trivial, especially under the desideratum that the active spaces comprise a few electrons and orbitals.
We observed that occupied localized DFT orbitals can be reliably ranked based on their contribution to the difference between the DFT density of the adsorbate+surface complex and the sum of the individual DFT densities of adsorbate and surface, Eq.~\eqref{eq:chrg_den_diff}.
The ranking and selection of virtual DFT orbitals is more delicate and requires estimating their contribution to the active-space ground-state energy. In line with chemical knowledge~\cite{smith2017cheap, motta2020ground, motta2017towards, Yingjin2013}, we observed that natural orbitals lead to faster convergence of active-space ground-state energies than DFT orbitals, as they can capture anti-bonding virtual orbitals as opposed to the Rydberg continuum~\cite{helgaker2014molecular}.
In this work, we used CCSD calculations to construct natural orbitals, which are expensive for large systems. However, this issue can be mitigated by pre-screening virtual orbitals based on the density difference Eq.~\eqref{eq:chrg_den_diff} and/or resorting to less expensive MBPT2 (many-body second-order perturbation theory) calculations for dynamically correlated systems.

\paragraph{Active-space simulations:} In the solution of the Schr\"{o}dinger equation for the active space, we built upon recent work on the adaptation of quantum algorithms to crystalline solids \cite{yoshioka2020variational,Sokolov2020ooUCCSD, Mizukami2020,manrique2021momentumspace, yamamoto2021quantum,Yoshioka2021} and the design of variational ansatzes~\cite{QCC2018,eddins2022doubling} suited for near-term hardware.
An important contribution of our study is the introduction of a systematic and automated circuit simplification method based on iterative Clifford transformations. We tested the proposed circuit reduction technique focusing on the QCC ansatz and observed systematic and substantial reductions in the required number of CNOT gates and circuit depth, which stands to benefit simulations, especially on near-term quantum hardware.
Although investigated for the QCC ansatz, the circuit reduction technique proposed here is general, as it applies to any situation described by Eq.~\eqref{eq:Pauli_exponential}.

\paragraph{Applications and Perspective:}
Here, we demonstrated the proposed workflow using a step in the corrosion reaction of magnesium by water as an application. Previous studies have computed electronic and thermodynamic parameters using DFT~\cite{williams2015first, williams2016modeling, wurger2020first, yuwono2019aqueous, Yuwono2016, Limer2017}. Our study is a step towards the refinement of these calculations by (i) selecting a chemically meaningful and systematically improvable active space through an automatic, cost-effective procedure, and (ii) employing many-body methods in combination with an exact, automated, and general-purpose circuit reduction technique to simulate the active space.

The workflow proposed here is a natural choice in studying the adsorption and splitting of water onto a metal surface, which is an example of a broader class of reactions. These reactions include heterogeneous catalysis and atmospheric corrosion of substrates (e.g., surfaces made of transition metals and/or containing defects) by adsorbates (e.g., O$_2$), and involve bond breaking/formation in spatially localized regions, making them amenable to description through embedding and quantum computing active-space simulations.

Finally, the proposed workflow is valuable for both near- and long-term quantum computers. In the near term, it enables studies of complex chemical phenomena on noisy quantum devices by selecting relevant degrees of freedom and reducing quantum resource requirements. In the long term, it can support the study of systems with strong and spatially local electronic correlation, where traditional methods become less reliable~\cite{Wecker2014, Goings2022, Elfving2020}, using sophisticated algorithms like quantum phase estimation. Therefore, this workflow indicates a promising direction in the search for advantageous applications of quantum simulation algorithms. 

\section{Methods}
\label{sec:Methods}

\subsection{Geometry Optimization}

We performed geometry optimizations using plane-wave bases with Quantum ESPRESSO~\cite{QE-2009} (QE) v6.3. We modeled the \textit{hcp} Mg(0001) surface using a slab of 4 Mg layers with 16 atoms per layer, having a thickness of 
7.8 \r{A} having 10.5 \r{A} of vacuum separation between periodic images. Mg atoms in the 2 bottom layers were kept fixed in their positions, and those in the 2 top layers were allowed to relax. 

We performed calculations with the Perdew-Wang 91 DFT functional~\cite{burke1998derivation}, the scalar-relativistic Vanderbilt ultra-soft pseudo-potential~\cite{vanderbilt1990soft}, and a 30/360 Ry cutoff for wavefunction/density plane-wave expansion. 
We tested Brillouin zone convergence, for the optimized reactant and product geometries, using meshes of up to \bzmesh{8}{8}{1} $\vett{k}$ points (see the SI). Results from \bzmesh{4}{4}{1} and \bzmesh{8}{8}{1} meshes are within 1.6 kcal/mol of each other.

\subsection{DFT calculations in a Gaussian basis}

At the optimized geometries described above, we performed DFT calculations using a single-particle basis of translational-symmetry-adapted linear combinations of Gaussian atomic orbitals (AOs),
\begin{equation}
\label{eq:periodic_basis}
\varphi_{ \vett{k},p}( \vett{r} ) = \sum_{ \vett{T} } e^{ i \vett{k} \cdot \vett{T} } \chi_p( \vett{r} - \vett{T} )
\;.
\end{equation}
Here, $\vett{T} = \sum_{i=1}^3 T_i \vett{a}_i$ is a lattice translation vector, $\vett{k} = \sum_{i=1}^3 k_i \vett{b}_i$ is a momentum vector in the first Brillouin zone of the lattice, and $\chi_p$ is an orbital from a Gaussian basis set.
The summation over $\vett{T}$ leads to a basis of translational-symmetry-adapted orbitals~\cite{dovesi2005ab,mcclain2017gaussian}.

We performed DFT calculations with the PySCF package~\cite{sun2018pyscf,sun2020recent}. We used the GTH-DZV basis set~\cite{goedecker1996separable}, the associated Goedecker-Teter-Hutter (GTH) pseudo-potential~\cite{goedecker1996separable},
and the Perdew-Burke-Ernzerhof (PBE) functional~\cite{PBE1996}.

In the basis \eqref{eq:periodic_basis}, the Born-Oppenheimer Hamiltonian takes the form~\cite{vandevondele2005quickstep}
\begin{equation}
\label{eq:periodic_hamiltonian}
\hat{H} = E_0 + \sum_{ \substack{\vett{k} \\ pr \\ \sigma} } h_{pr}(\vett{k}) \crt{\vett{k}p \sigma} \dst{\vett{k}r \sigma}
+ \sum^*_{ \substack{\vett{k}_p \vett{k}_r \vett{k}_q \vett{k}_s \\ prqs \\ \sigma\tau} } 
\frac{ ( \vett{k}_p p, \vett{k}_r r | \vett{k}_q q , \vett{k}_s s ) }{2} \crt{\vett{k}_p p \sigma} \crt{\vett{k}_q q \tau} \dst{\vett{k}_s s \tau} \dst{\vett{k}_r r \sigma}
\;,
\end{equation}
where the $*$ symbol indicates crystal momentum conservation, i.e. $\vett{k}_p + \vett{k}_q - \vett{k}_r - \vett{k}_s = \vett{G}$, where $\vett{G}$ is a reciprocal lattice vector. Here, we approximated the electron-electron interaction with density fitting using a Weigend auxiliary basis.

\subsection{Active-Space Selection}

\paragraph{Orbital localization:} We localized Kohn-Sham orbitals using Pipek-Mezey method~\cite{PipekMezey1989,PM2014,QSunLoc2014} based on a Mulliken population analysis with meta-L\"owdin orbitals.~\cite{QSunLoc2014,PM2014}. We conducted separate localization for occupied and virtual orbitals, at each $\vett{k}$ point individually.

\paragraph{Density difference:} We computed the electronic density difference, Eq.~\eqref{eq:chrg_den_diff}, at DFT level. This quantity is shown in the SI for the reactant and product. For each localized DFT orbital $\psi_\ell(\vett{x})$, we computed the overlap function
\begin{equation}
\mathcal{O}_\ell(\vett{x}) = 
\sqrt{ \rho_{\mathrm{DD}}(\vett{x}) }
\,
| \psi_\ell(\vett{x}) |
\label{eq:ovp}
\end{equation}
and the integrated overlap
\begin{equation}
\tilde{\mathcal{O}}_\ell[\eta] = \int d\vett{x} \quad \mathcal{O}_\ell(\vett{x}) H(\mathcal{O}_\ell(\vett{x})-\eta)
\;.
\label{eq:ovp_thrsh}
\end{equation}
In Eq.~\eqref{eq:ovp_thrsh}, the parameter $\eta$ is a threshold used to truncate the tails of $\mathcal{O}_\ell(\vett{x})$ and $H(x)$ is Heaviside's step function.
For small $\eta$, due to such tails, physically irrelevant orbitals delocalized across the metallic surface have artificially high integrated overlaps.
For large $\eta$, all orbitals (including physically relevant ones) have zero integrated overlap.
However, there is a broad region of values of $\eta$, that we identified by a simple scan, that leads to a stable ordering of orbitals.
We then used integrated overlap $\tilde{\mathcal{O}}_\ell[\eta]$ to rank the occupied localized DFT orbitals.

\subsection{Variational ansatzes for active-space simulations}

\paragraph{Unitary CCSD:} This ansatz is obtained by applying the exponential of the anti-Hermitian operator $T-T^{\dagger}$ to the Hartree-Fock state. $T$ is a linear combination of single and double excitations from occupied (indexed as $j,k$) to virtual spin-orbitals (indexed as $b,c$). More specifically,
\begin{eqnarray}
\label{eq:quccsd}
|\psi_{\mathrm{UCCSD}}(\boldsymbol{\theta})\rangle 
&=& 
e^{T-T^{\dagger}}|\psi_{\mathrm{HF}}\rangle \\
T
&=& 
T_{1} + T_{2} \\
T_{1}
&=&
\sum_{aj} (\theta^{R}_{aj}+i\theta^{I}_{aj}) \crt{a} \dst{j} \\
T_{2} 
&=& 
\sum_{ajbk}
(\theta^{R}_{ajbk}+i\theta^{I}_{ajbk})
\crt{a} \crt{b} \dst{k} \dst{j}
\;,
\end{eqnarray}
where $\crt{a}$/$\dst{j}$ creates/destroys an electron at spin-orbital $a$/$j$,
and the coefficients $\boldsymbol{\theta} = \{\theta^{R}_{aj}$, $\theta^{I}_{aj}$, $\theta^{R}_{ajbk}$, $\theta^{I}_{ajbk}\}$ are variational parameters.
At the $\vett{\Gamma}$ point, where the Hamiltonian Eq.~\eqref{eq:periodic_hamiltonian} is time-reversal-symmetric,
electronic eigenfunctions are real-valued and coefficients $\theta^{I}$ can be forced to zero. Away from the $\vett{\Gamma}$ point, this is no longer true.
To implement the ansatz Eq.~\eqref{eq:quccsd} on a gate-based quantum computer, we mapped the fermionic operator $T-T^{\dagger}$ onto a linear combination of Pauli operators. Then, we approximated $\exp(T-T^\dagger)$ with a Trotter-Suzuki approximation or other product formulas, yielding the Trotterized form of UCCSD, known as qUCCSD~\cite{Barkoutsos2018}.
We implemented the qUCCSD using Qiskit~\cite{Qiskit}, with appropriate modifications to include the coefficients $\theta^{I}$.

\paragraph{Qubit coupled-cluster:} the QCC ansatz~\cite{QCC2018,QCC2020} is defined by sequentially applying exponentials of Pauli strings from an ordered set $P_1 \dots P_m$ to the Hartree-Fock state as in Eq.~\eqref{eq:QCC1}.
The Pauli operators $P_{k}$ are typically~\cite{QCC2018, QCC2020} ranked based on the value of the energy gradient $g(P) = |\langle \psi_{\mathrm{HF}} | [H,P] | \psi_{\mathrm{HF}} \rangle|$. The number $m$ is determined based on the convergence of the QCC energy or the budget of the simulator/hardware at hand.
In this study, for simplicity, we elected to choose the operators $P_k$ based on the coefficients of the CASCI wavefunction $| \psi_{\mathrm{CASCI}} \rangle = \sum_{c} v_c | \psi_c \rangle$. More specifically, we sorted the configurations $\psi_c$ in decreasing order of $|v_c|$ and retained the top $m/2$ configurations.
For each such configuration, we constructed two Pauli operators $P_c, Q_c$ such that $P_c | \psi_{\mathrm{HF}} \rangle = | \psi_c \rangle$ and $Q_c | \psi_{\mathrm{HF}} \rangle = i |\psi_c \rangle$, and introduced them in the pool of QCC Pauli operators.
We implemented the QCC ansatz using Qiskit and optimized it with the $\text{L\_BFGS\_B}$~\cite{LBFGSB2003} algorithm on noiseless classical simulators.

\paragraph{Entanglement Forging:} this algorithm~\cite{eddins2022doubling} writes a wavefunction $\Psi$ of a bipartite quantum system $A+B$ through a Schmidt decomposition,
\begin{equation}
|\Psi_{\mathrm{A}+\mathrm{B}}\rangle = \sum_{i=1}^R \lambda_i 
U | x_i \rangle \otimes V | y_i \rangle
\;.
\label{eq:EF_wf}
\end{equation}
In Eq.~\eqref{eq:EF_wf}, $U$ and $V$ are unitary matrices, $\lambda_i$ are Schmidt coefficients, and $x_i,y_i$ are computational basis states (or bitstrings). The number $R$ is called the Schmidt rank of $\Psi_{\mathrm{A}+\mathrm{B}}$, and depends on the entanglement across $A$ and $B$.
Operators like the Hamiltonian are written as linear combinations $H = \sum_{a,b =1}^{h}w_{a,b} P_{a}\otimes P_{b}$ of Pauli operators acting on $A$ and $B$ individually, and its expectation value over $\Psi_{\mathrm{A}+\mathrm{B}}$ is written as
\begin{equation}
E = 
\sum_{\substack{a,b \\ i,j}}
w_{a,b}
\lambda_{i}\lambda_{j}
\langle x_j | U^{\dagger} P_{a} U | x_i \rangle
\langle y_j | V^{\dagger} P_{b} V | y_i \rangle
\;.
\label{eq:EF_eq}
\end{equation}
The cost of evaluating $E$ scales as $O(R^2 h)$. The terms in Eq.~\eqref{eq:EF_eq} can be evaluated on a quantum computer, using half of the qubits required to store $\Psi_{\mathrm{A}+\mathrm{B}}$ when $A$ and $B$ have equal sizes.
In this work, we used EF as a variational ansatz for VQE calculations. We chose bitstrings $x_i=y_i$ and unitaries $U=V$ as detailed in the SI, and optimized parameters with the Simultaneous Perturbation Stochastic Approximation (SPSA)~\cite{spall1992multivariate} method.

\subsection{Hardware Experiments}

We performed hardware experiments using multiple IBM Quantum devices accessed via cloud, specifically, $\text{ibmq}\_\text{lima}$, $\text{ibmq}\_\text{guadalupe}$, $\text{ibmq}\_\text{toronto}$, $\text{ibmq}\_\text{casablanca}$, $\text{ibm}\_\text{perth}$,
$\text{ibm}\_\text{brisbane}$,
$\text{ibm}\_\text{sherbrooke}$,
and $\text{ibm}\_\text{lagos}$. 

For the HONO-LUNO active space (Fig.~\ref{fig:qc_algos}, left panel) we performed a full VQE calculation for each $\vett{k}$ point using the COBYLA optimizer~\cite{Powell1994}. We used readout error mitigation as well as gate-based zero-noise extrapolation~\cite{Rice2021} to mitigate hardware noise.
Furthermore, for each VQE calculation, we carried out 5 independent trials yielding 5 sets of optimized parameter configurations.  For each such configuration, we ran a two-point gate-based zero-noise extrapolation and averaged the extrapolated results.

For 10-orbital active spaces (Fig.~\ref{fig:qc_algos}, right panel) we used optimized parameters from classical noiseless simulations to compute the VQE energy for QCC with 2 and 5 Pauli strings. 5 independent hardware experiments were performed for the reactant and product in each case, and the corresponding standard deviation in hardware data is provided as error bars for both active spaces.
Readout error mitigation was used along with gate-based zero-noise extrapolation as needed. The circuits produced by the reduction technique are shown in Supplementary Figures~12 and 13.

\section{Data availability}
The numerical and quantum hardware data generated and analyzed during the current study will be made available from the corresponding authors upon reasonable request.

\section{Code availability}
The code used for generating data during the current study will be made available from the corresponding author upon reasonable request.

\section{Acknowledgments}
We thank John Lowell, Brittan Farmer, and Benjamin Koltenbah for their helpful comments and discussions.

\section{Competing interests}

The Authors declare no competing financial or non-financial interests.

\section{Author Contributions}
K.W., J.E.R., B.A.J., M.M., T.P.G., N.N., R.J.T., and M.K. developed the proposal and designed the study. T.P.G. lead the technical coordination of the project. T.P.G., M.M., N.N., T.N.F., P.Kl.B., T.S. contributed towards code development and performed the computations. T.P.G, N.N., M.M., and J.E.R. worked on the development of the circuit reduction technique.  N.N. and T.P.G. performed experiments on quantum hardware. All authors contributed to analyzing the data. T.P.G., M.M., J.E.R., T.N.F., P.Kl.B., B.A.J. contributed towards writing the manuscript. All authors contributed to the improvement of the manuscript.

\end{document}


\title{Supplementary Information for\\ ``Quantum Computation of Reactions on Surfaces Using Local Embedding''}

\author{Tanvi P. Gujarati}
\affiliation{IBM Quantum, Almaden Research Center, San Jose, California 95120, USA}
\email{tgujarati@ibm.com; mario.motta@ibm.com}
\author{Mario Motta}
\affiliation{IBM Quantum, Almaden Research Center, San Jose, California 95120, USA}
\author{Triet Nguyen Friedhoff}
\affiliation{IBM Quantum, T. J. Watson Research Center, Yorktown Heights, New York 10598, USA}
\author{Julia E. Rice}
\affiliation{IBM Quantum, Almaden Research Center, San Jose, California 95120, USA}
\author{Nam Nguyen}
\affiliation{Integrated Vehicle Systems, Applied Mathematics, Boeing Research \& Technology, Huntington Beach, CA, 92647}
\email{nam.h.nguyen5@boeing.com}
\author{Panagiotis Kl. Barkoutsos}
\affiliation{IBM Quantum, IBM Research – Zurich, 8803 Rüschlikon, Switzerland}
\author{Richard J. Thompson}
\affiliation{Integrated Vehicle Systems, Applied Mathematics,  Boeing Research \& Technology, Huntsville, AL, 35824}
\author{Tyler Smith}
\affiliation{Integrated Vehicle Systems, Applied Mathematics,  Boeing Research \& Technology, Orlando, FL, 32826}
\author{Marna Kagele}
\affiliation{Tech Vis and Integration, Global Technology, Boeing Research \& Technology, Tukwila, WA 98108}
\author{Mark Brei}
\affiliation{BSC Analytics, Chemical Technology, North Charleston, SC, 29456}
\author{Barbara A. Jones}
\affiliation{IBM Quantum, Almaden Research Center, San Jose, California 95120, USA}
\author{Kristen Williams}
\affiliation{Integrated Vehicle Systems, Applied Mathematics,  Boeing Research \& Technology, Huntsville, AL, 35824}

\maketitle

\section{Supplementary Methods}

\subsection*{Corrosion Reaction}
The overall corrosion reaction \cite{williams2016modeling_si, esmaily2017fundamentals_si} in aqueous environments is given as:

\begin{equation}
\ce{Mg + 2H2O <=>> \text{\ce{Mg(OH)2}} + H2}
\label{eq:Mg_cor_main_si}
\end{equation}
Eq. (\ref{eq:Mg_cor_main_si}) contains two partial equations corresponding to the oxidation of magnesium and reduction of water:
\begin{equation}
\ce{Mg <=>> \text{\ce{Mg^2+}} + 2 e^-} (anodic)
\label{eqn:anodic}
\end{equation}
\begin{equation}
\ce{2H2O + 2e^- <=>>\text{\ce{H2}} + 2OH^-} (cathodic)
\label{eqn:cathodic}
\end{equation}
Williams et al \cite{williams2016modeling_si} proposed a detailed reaction scheme connecting the steps of initial water dissociation on Mg surface with the final step of $\ce{H_{2}}$ evolution via Tafel mechanism \cite{Sharifi2013_si, williams2016modeling_si} in the presence of adsorbed OH$_\text{ads}$ and H$_\text{ads}$ species using modeling based on DFT. The suggested reaction mechanism was shown to be a concerted reaction involving multiple water molecules. The first reaction studied in the process was the splitting of a single \ce{H_{2}O} molecule creating adsorbed H$_\text{ads}$ and OH$_\text{ads}$ moiety:
\begin{equation}
\ce{Mg + H_{2}O -> Mg(OH_{ads})(H_{ads}) }
\label{eq:Mg_singleH2O_si}
\end{equation}
In the presence of multiple water molecules, the proposed reaction mechanism consists of three steps:
\begin{eqnarray}
\label{eq:Mg_concerted1}
\ce{Mg + 2H_{2}O & -> & Mg(OH_{ads})(H_{ads}) + H_{2}O}\\
\ce{Mg(OH_{ads})(H_{ads}) + H_{2}O & -> & Mg(OH_{ads})_{2}(H_{ads})_{2}}\\
\ce{Mg(OH_{ads})_{2}(H_{ads})_{2} & -> & Mg(OH_{ads})_{2} + H_{2}(g)}
\label{eq:Mg_concerted3}
\end{eqnarray}

\subsection*{Geometry Optimization}
Differences between product and reactant ground-state energies based on Density Functional Theory (DFT) are shown, as a function of the total number of $\vett{k}$ points in Monkhorst-Pack meshes of size $y\times y\times 1$ with $y \in \{1,2,4,6,8\}$, in Supplementary Figure {\ref{fig:QE_conv}}. These calculations were performed using Quantum Espresso, a plane-wave basis set, and full $\vett{k}$-point sampling.

\begin{figure}[h!]
	\centering
	\includegraphics[width=0.5\textwidth]{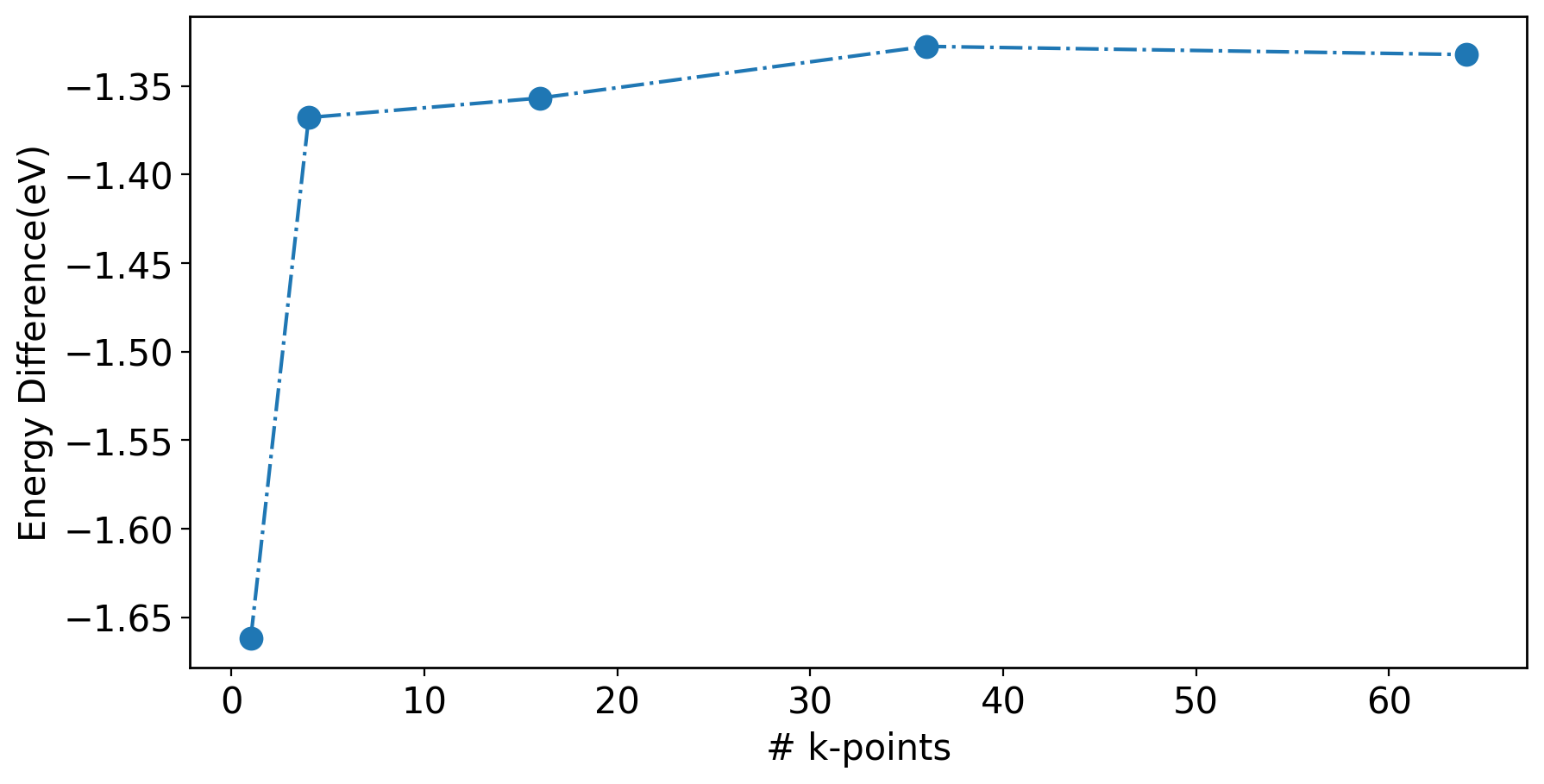}
	\caption{Energy difference between the product and reactant as a function of the number of $\vett{k}$-points}
	\label{fig:QE_conv}
\end{figure}

\subsection*{Active-Space Selection}

Electronic Density Differences (DD) for the reactant and product systems, obtained from DFT calculations at the $\vett{\Gamma}$ point using Gaussian basis sets, are shown in Supplementary Figure \ref{fig:den_diff}. DDs are plotted with VESTA~\cite{VESTA2011_si}.

The top five orbitals selected from the ordered set of occupied orbitals produced after sorting based on electronic DD for the reactant and product at the $\vett{\Gamma}$ point are shown in Supplementary Figure \ref{fig:occupied_orbitals}. All the occupied orbitals are localized around the surface moieties. A similar trend is observed for other $\vett{k}$ points as well.

\begin{figure}[h!]
\centering
\includegraphics[width=\textwidth, trim={0cm 12cm 9cm 0cm},clip]{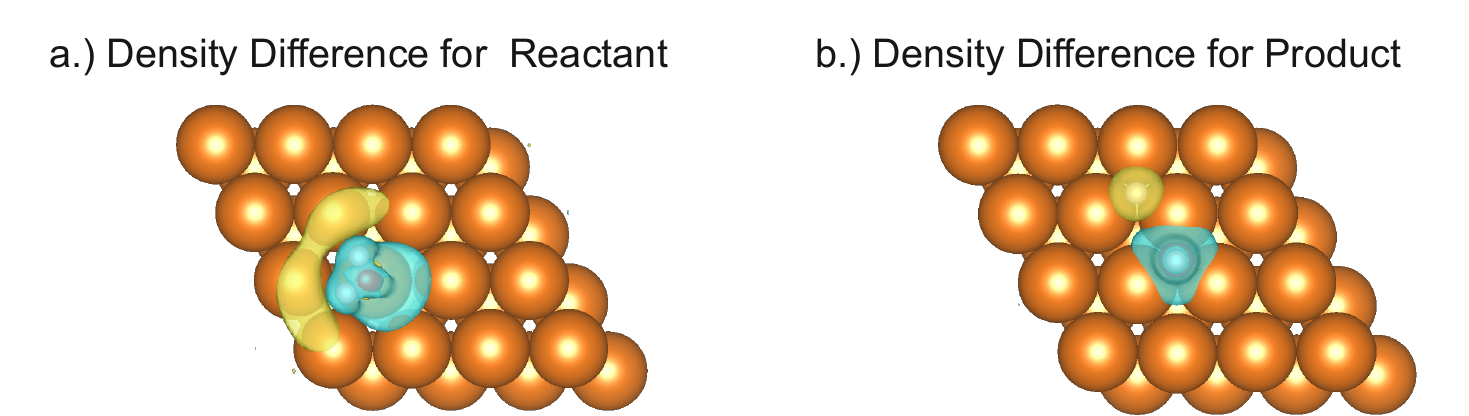}
\caption{Electronic DD for the reactant and product systems as seen along the $z$ axis.}
\label{fig:den_diff}
\end{figure}
\begin{figure}[h!]
\centering
\includegraphics[width=\textwidth, trim={0cm 2cm 0cm 0cm},clip]{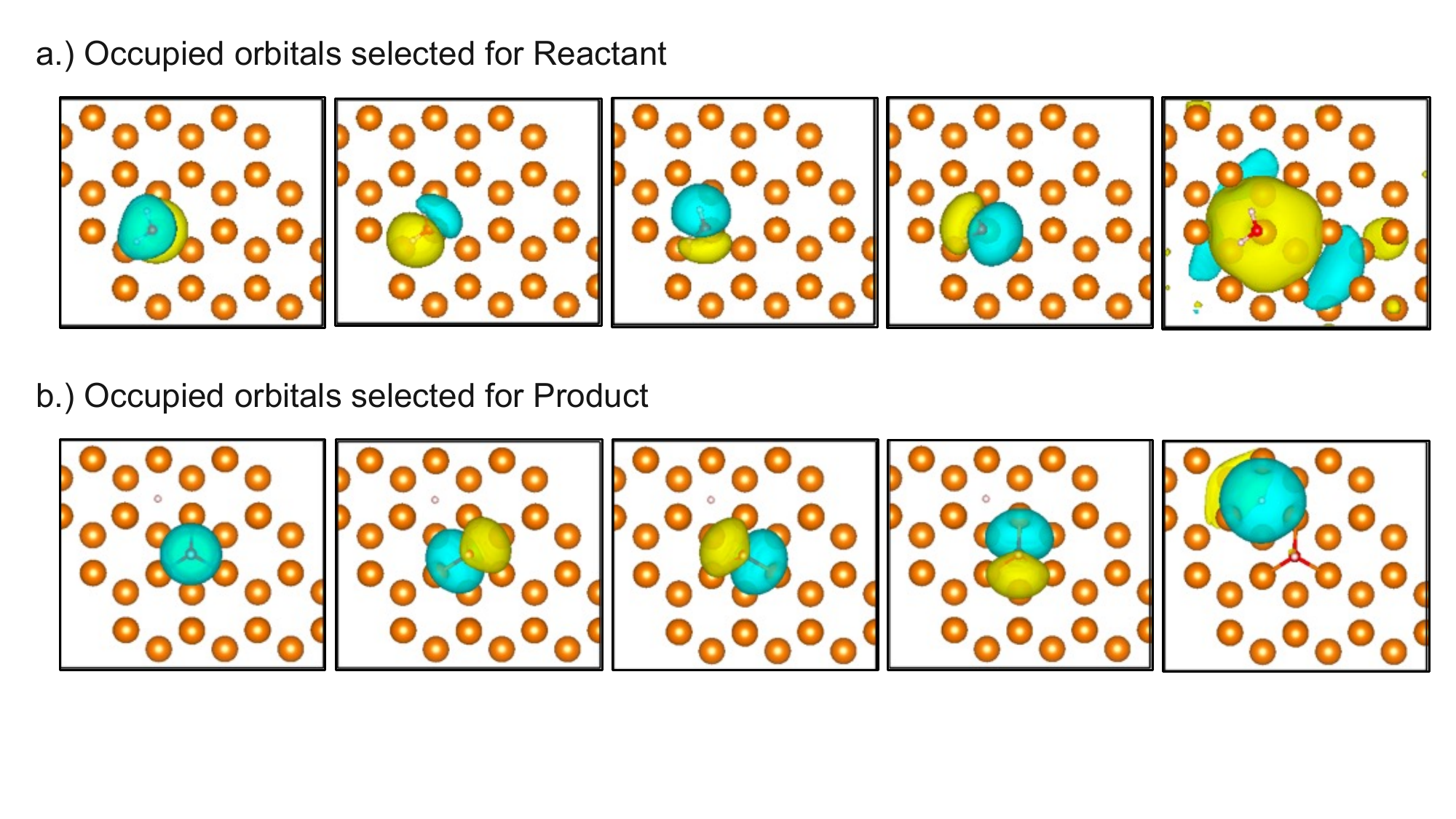}
\caption{Selected occupied orbitals for the reactant and product systems based on electronic DD at the $\vett{\Gamma}$ point.}
\label{fig:occupied_orbitals}
\end{figure}

\clearpage
    
\subsection*{VQE results with Qubit Coupled Cluster ansatz}
VQE results obtained using the Qubit Coupled Cluster (QCC) ansatz with an increasing number of Pauli strings for all $\vett{k}$ points of a $4\times4\times1$ Monkhorst-Pack grid are shown in Supplementary Figure \ref{fig:QCC_ediff}-\ref{fig:QCC_angmom}. Each panel of Supplementary Figure \ref{fig:QCC_ediff} shows the energy difference between the product and reactant minimum energy states in active space sizes increasing from 2 (top panel) to 10 (bottom panel) created using the electronic density difference + natural orbitals (DD+NO) method of active space selection. Similarly, Supplementary Figure \ref{fig:QCC_pnum} shows the number of particles for the reactant (on left) and product (on right) with increasing active space sizes going from 2 (top) to 10 (bottom). A maximum deviation of $~10^{-3}$ from the expected value is observed from these results. Results for other properties such as the $z$-component of the spin ($S_{z}$ in units of $\hbar$)  and total spin ($S^{2}$ in units of $\hbar^2$) are shown in Supplementary Figure \ref{fig:QCC_mag} and Supplementary Figure \ref{fig:QCC_angmom} respectively. The discrepancy in the total spin gets worse as the active space size increases. After the addition of a sufficient number of Pauli strings in the ansatz, we expect the discrepancy to disappear.

\begin{figure}[h]
\centering
\includegraphics[width=\textwidth,height=0.8\textheight,keepaspectratio]{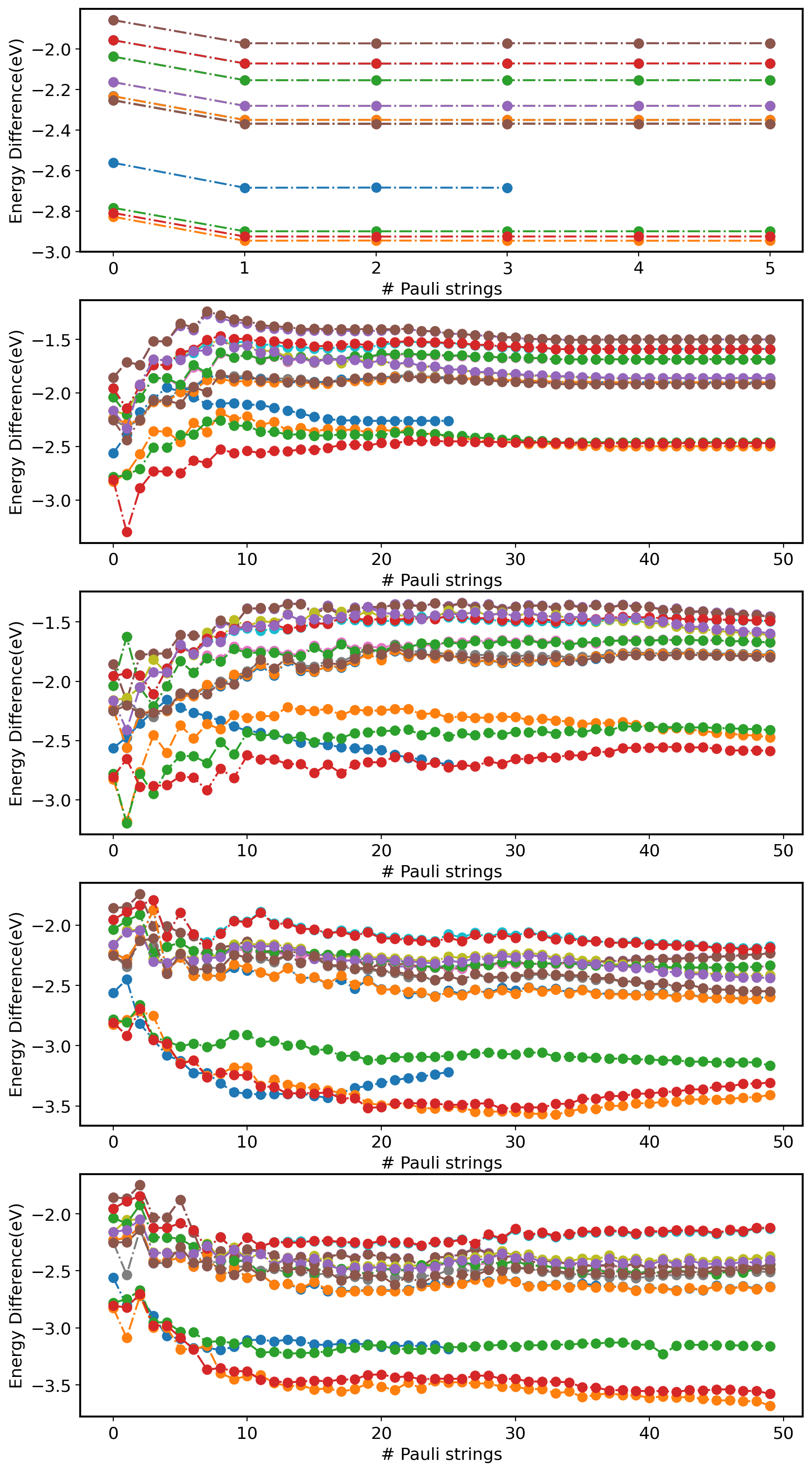}
\caption{Energy difference between the product and reactant as function of number of Pauli strings added to the QCC ansatz. Top panel corresponds to active space of HONO-LUNO orbitals. The active space size increases by 2 with each panel going downwards until the bottom panel with active space of 10 natural orbitals. Within each panel 16 curves represent the 16 $\vett{k}$-points. One curve in each panel needs only half the number of Pauli strings since it describes higher symmetry $\vett{\Gamma}$-point.}
\label{fig:QCC_ediff}
\end{figure}

\clearpage
\begin{figure}[h]
\centering
\includegraphics[width=\textwidth,height=0.9\textheight,keepaspectratio]{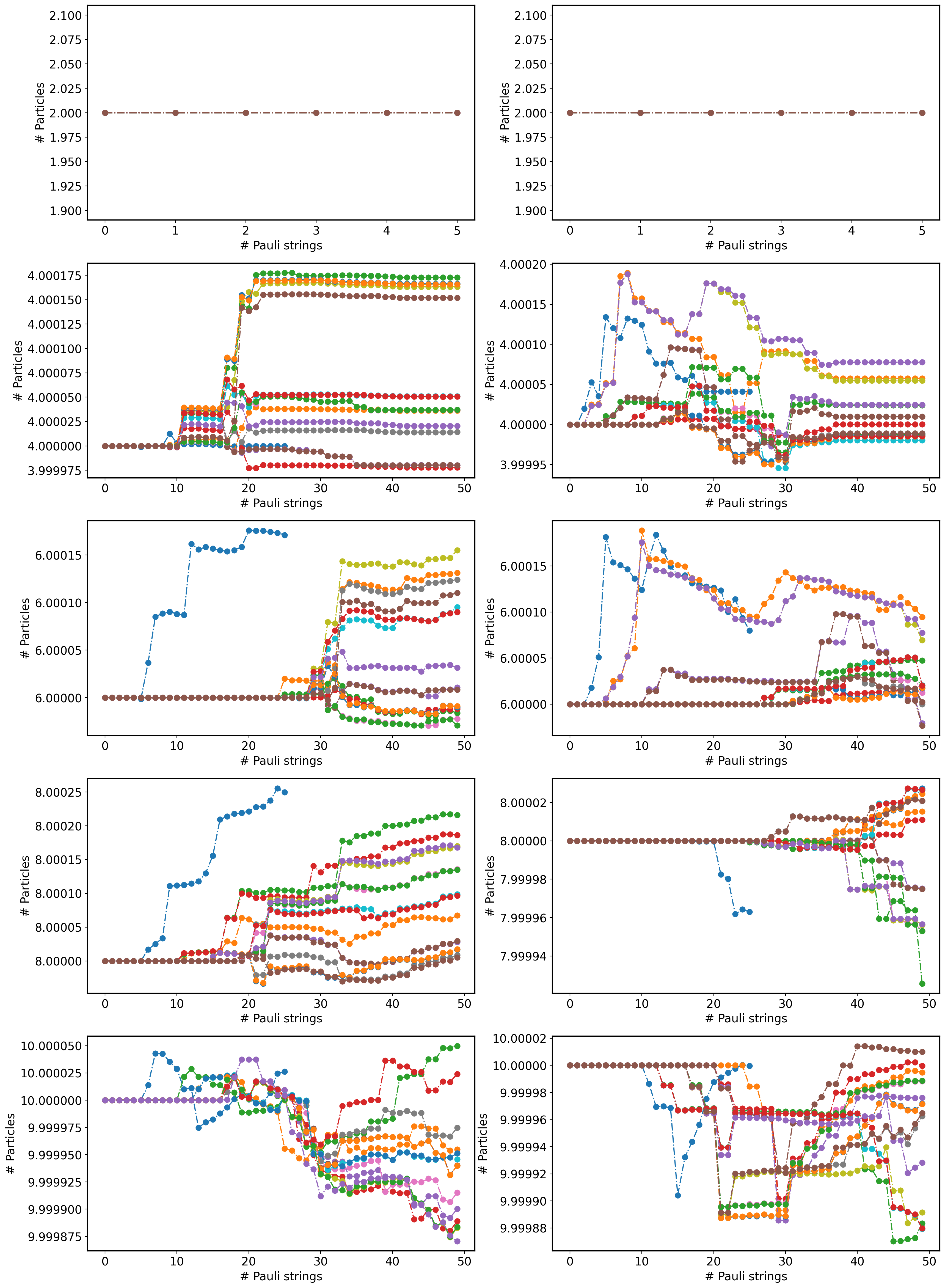}
\caption{Number of particles for the reactant (left) and product (right) as function of number of Pauli strings added to the QCC ansatz. Top panel corresponds to active space of HONO-LUNO orbitals. The active space size increases by 2 with each panel going downwards. All 16 points are shown in each panel.}
\label{fig:QCC_pnum}
\end{figure}

\newpage
\begin{figure}[h]
\centering
\includegraphics[width=\textwidth,height=0.9\textheight,keepaspectratio]{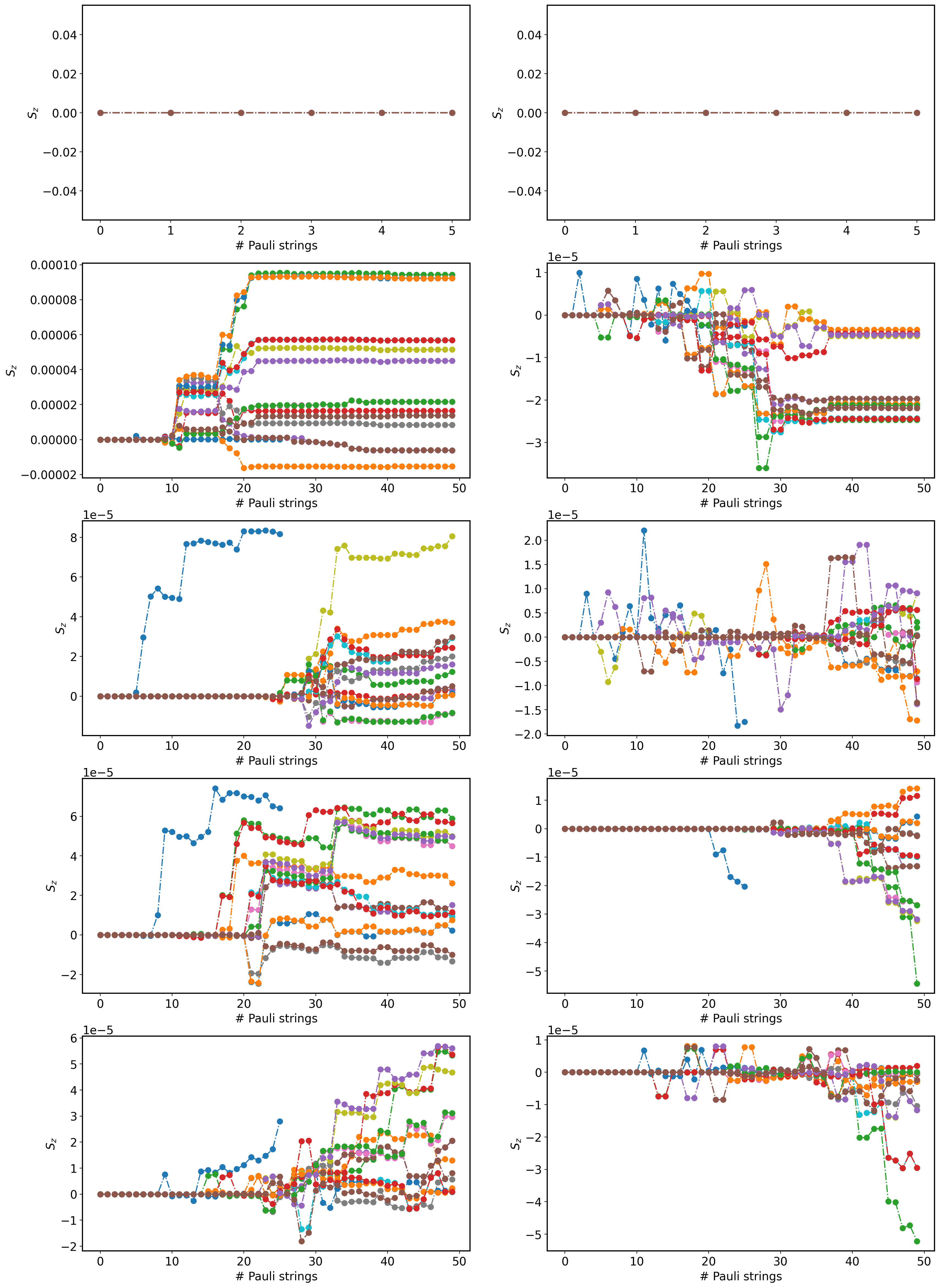}
\caption{The z-component of the spin $S_{z}$ for the reactant (left) and product (right) as function of number of Pauli strings added to the QCC ansatz. Top panel corresponds to active space of HONO-LUNO orbitals. The active space size increases by 2 with each panel going downwards. All 16 points are shown in each panel.}
\label{fig:QCC_mag}
\end{figure}

\newpage
\begin{figure}[h]
\centering
\includegraphics[width=\textwidth,height=0.9\textheight,keepaspectratio]{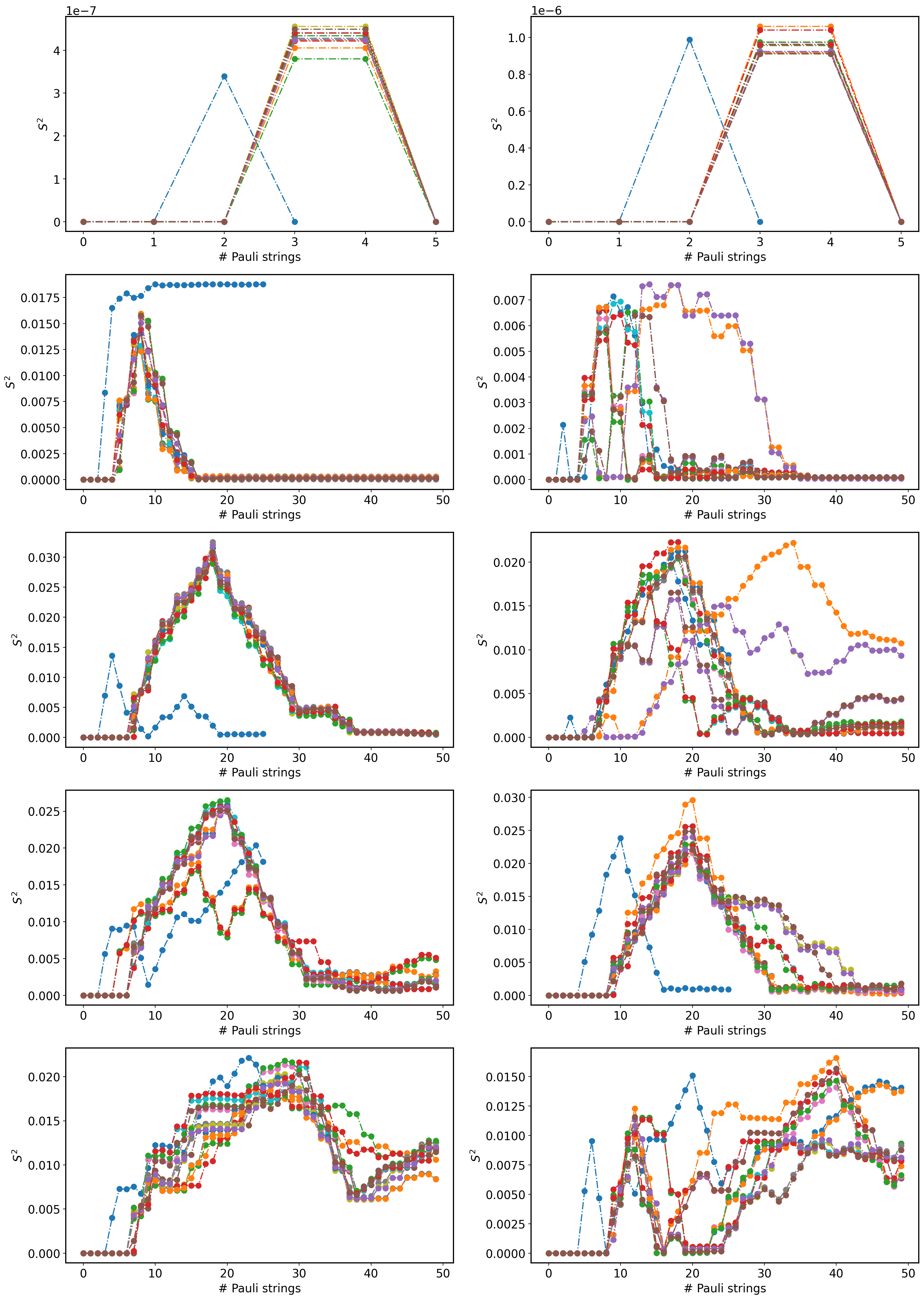}
\caption{The total spin $S^{2}$ for the reactant (left) and product (right) as function of number of Pauli strings added to the QCC ansatz. Top panel corresponds to active space of HONO-LUNO orbitals. The active space size increases by 2 with each panel going downwards. All 16 points are shown in each panel.}
\label{fig:QCC_angmom}
\end{figure}

\clearpage
\subsection*{Entanglement Forging}
\label{sec:S5}
In Supplementary Figure \ref{fig:EF_sv} the singular values obtained from a Singular Value Decomposition of the CASCI (Complete Active Space Configuration Interaction) wavefunction by partitioning along the spin-up and spin-down sectors for each active space constructed from the DD+NO method for the reactant and product is shown. The results were evaluated at the $\vett{\Gamma}$-point. Top panel is for the HONO-LUNO orbitals and the active space size increases by 2 orbitals with each panel going downwards up to 10 natural orbitals. The singular values fall off quickly in all cases suggesting that the spin-up and spin-down subsystems in the ground state within each active space are weakly entangled.

\begin{figure}[h!]
	\centering
	\includegraphics[width=\textwidth, height = 0.9\textheight, keepaspectratio]{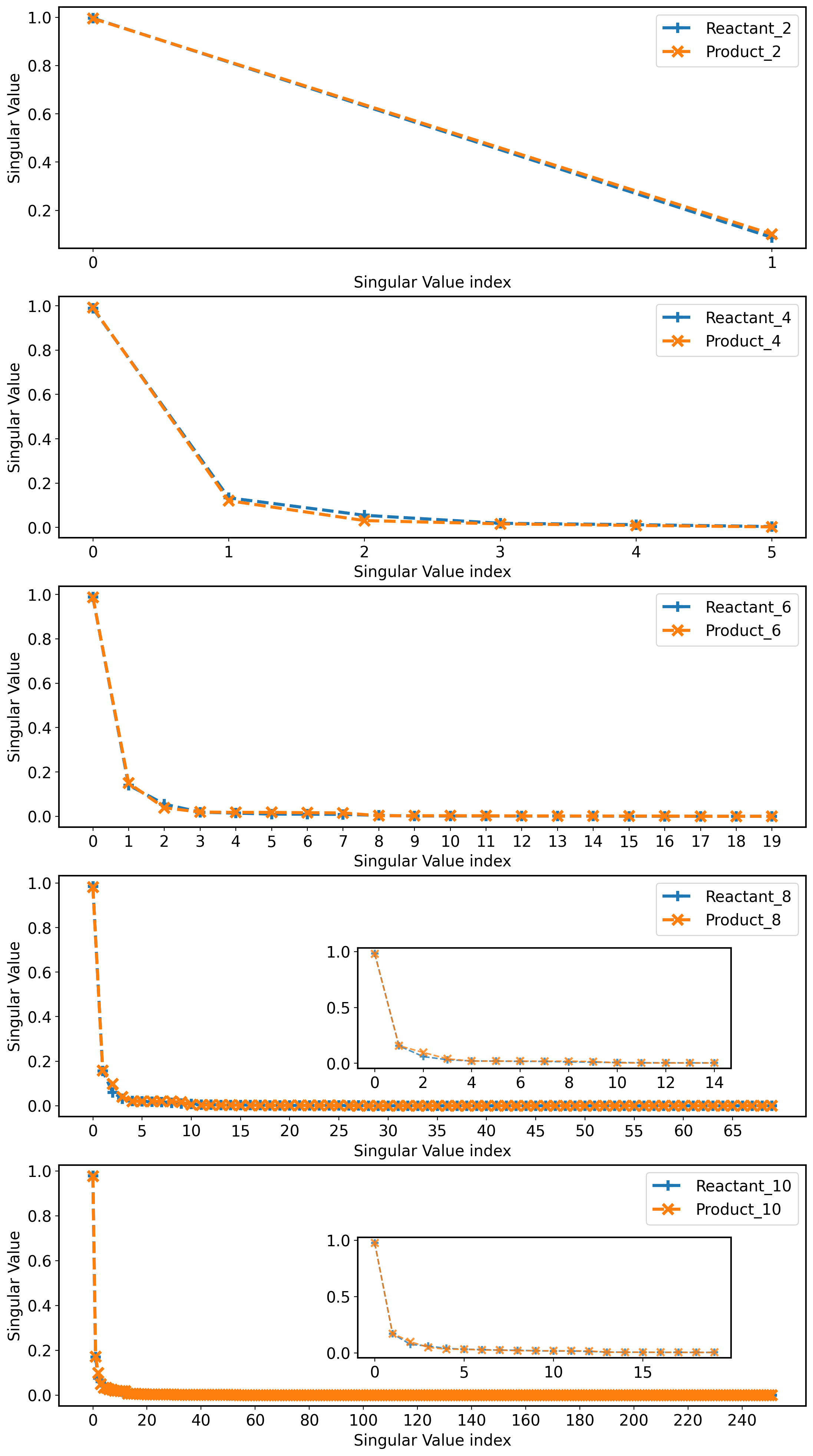}
	\caption{Singular values of the CASCI wavefunction for the reactant and product in increasing active space sizes going from 2 (top) to 10 (bottom) at the $\vett{\Gamma}$-point. Inset in bottom two panels highlights the zoomed in results for small singular value indices.}
	\label{fig:EF_sv}
\end{figure}

In the Entanglement Forging (EF) algorithm \cite{eddins2022doubling_si}, $|x \rangle_A$ and $|x \rangle_B$ represent the N-qubit bit-string states or electronic configurations that describe a state of a bipartite ($N+N$)-qubit system $| \Psi_{A+B} \rangle $. Since each bit-string describes electron occupations for spin-up and spin-down parts separately, the length of each bit-string corresponds to the number of active orbitals. EF calculations at the $\vett{\Gamma}$ point for active-space sizes ranging from 2 to 10 active natural orbitals were performed. Table \ref{EF_gamma_1} (for reactant) and Table \ref{EF_gamma_23} (for product) provide the details of the bit-strings used as well as the hop gate configuration selected for the $U$ and $V$ unitary matrices. Each hop gate, $h(\varphi)$, acts according to the following matrix:

\begin{equation*}
h(\varphi) = 
\begin{bmatrix}
1 & 0 & 0 & 0 \\
0 & \cos(\varphi) & -\sin(\varphi) & 0 \\
0 & \sin(\varphi) &  \cos(\varphi) & 0 \\
0 & 0 & 0 & -1
\end{bmatrix}
\end{equation*}

While testing various ansatzes for EF calculations, we systematically increased the number of included bit strings from one to four, from a list of the most dominant bit strings obtained from the full Configuration Interaction (CI) wavefunction in Jordan-Wigner (JW) transformation. For both reactant and product systems, each calculation includes the Hartree-Fock state, e.g., [1,1,0,0] for the case of four active orbitals. The obtained EF energies for each systems are reported as well. While the EF energies for the reactant and product become increasingly worse when compared to the CASCI energies as the active space size increases, the values of $\Delta E$ between the product and reactant compare well with the expected values as shown in Supplementary Figure \ref{fig:EF_ediff}. The CASCI energies are reported in Table \ref{CASCI_reactant} for the reactant and in Table \ref{CASCI_product} for the product.

\begin{table}[h!]
\centering
\def\arraystretch{1.5}
\begin{tabular}{|p{2.5cm}|p{5cm}|p{5cm}|p{3cm}|}
\hline\hline
\# Act. Orb. & Bit Strings & Hop Gates & EF Energy (Ha)\\ \hline\hline
2 & [1,0] & [0,1] & -3981.02732 \\ 
  & [0,1] &       &             \\ \hline
4 & [1,1,0,0]  & [0,1],[2,3] & -3981.05119 \\ 
  & [0,1,1,0]  &             &             \\ 
  & [0,1,0,1]  &             &             \\
  & [1,0,1,0]  &   &     \\ \hline
6 & [1, 1, 1, 0, 0, 0] & [0,1], [1,2], [2,0], [3,4], [4,5], & -3981.05198 \\ 
 & [0, 1, 1, 0, 1, 0]  & [5,3] &  \\ 
 & [0, 1, 1, 1, 0, 0]  &  &  \\ 
 & [1, 0, 1, 1, 0, 0]  &  &  \\ \hline
8 & [1, 1, 1, 1, 0, 0, 0, 0]   & [3,4], [2,3], [4,5], [1,2], [5,6], & -3981.05609 \\ 
 & [0, 1, 1, 1, 0, 1, 0, 0] & [0,1], [6,7] &  \\ 
 & [1, 1, 0, 1, 0, 0, 1, 0] &  &  \\ 
 & [1, 0, 1, 1, 1, 0, 0, 0] &  & \\ \hline
10 & [1, 1, 1, 1, 1, 0, 0, 0, 0, 0]  & [0,1], [0,2], [1,2], [1,3], [2,3],  & -3981.04709 \\
 & [1, 1, 1, 1, 0, 1, 0, 0, 0, 0] & [2,4], [3,4], [5,6], [5,7], [6,7],  &  \\
 & [1, 0, 1, 1, 1, 0, 0, 1, 0, 0] & [6,8], [7,8], [7,9], [8,9] & \\
 & [1, 1, 1, 0, 1, 0, 0, 0, 0, 1] &  &  \\\hline \hline
\end{tabular}
\caption{Entanglement forging set up and results for the reactant at the $\vett{\Gamma}$-point in different active space sizes.}
\label{EF_gamma_1}
\end{table}

\begin{table}[h!]
\centering
\def\arraystretch{1.5}
\begin{tabular}{|p{2.5cm}|p{5cm}|p{5cm}|p{3cm}|}
\hline\hline
\# Act. Orb. & Bit Strings & Hop-Gates & EF Energy (Ha)\\ \hline\hline
2 & [1,0] & [0,1] & -3981.12598 \\ 
  & [0,1] &       &             \\ \hline
4 & [1,1,0,0]  & [0,1],[2,3] & -3981.13349 \\ 
  & [0,1,1,0]  &             &             \\ 
  & [0,1,0,1]  &             &             \\
  & [1,0,0,1]  &   &     \\ \hline
6 & [1, 1, 1, 0, 0, 0] & [0,1], [1,2], [2,0], [3,4], [4,5], & -3981.15112 \\ 
 & [0, 1, 1, 0, 1, 0]. & [5,3], [2,3] &  \\ 
 & [0, 1, 1, 1, 0, 0]  &  &  \\ 
 & [1, 0, 1, 0, 0, 1]  &  &  \\ \hline
8 & [1, 1, 1, 1, 0, 0, 0, 0]   & [0,1], [2,3], [4,5], [6,7]  & -3981.14932 \\ 
 & [0, 1, 1, 1, 0, 1, 0, 0] & &  \\ 
 & [1, 1, 0, 1, 0, 0, 1, 0] & &  \\ 
 & [1, 1, 1, 0, 1, 0, 0, 0] &  & \\ \hline
10 & [1, 1, 1, 1, 1, 0, 0, 0, 0, 0]  & [0,1], [0,2], [1,2], [1,3], [2,3],  & -3981.15164 \\
 & [1, 1, 1, 1, 0, 1, 0, 0, 0, 0] & [2,4], [3,4], [5,6], [5,7], [6,7],  &  \\
 & [1, 0, 1, 1, 1, 0, 1, 0, 0, 0] & [6,8], [7,8], [7,9], [8,9] & \\
 & [1, 1, 1, 0, 1, 0, 0, 1, 0, 0] &  &  \\\hline \hline
\end{tabular}
\caption{Entanglement forging set up and results for the product at the $\vett{\Gamma}$-point in different active space sizes.}
\label{EF_gamma_23}
\end{table}

\begin{figure}[h!]
	\centering
	\includegraphics[width=0.6\textwidth]{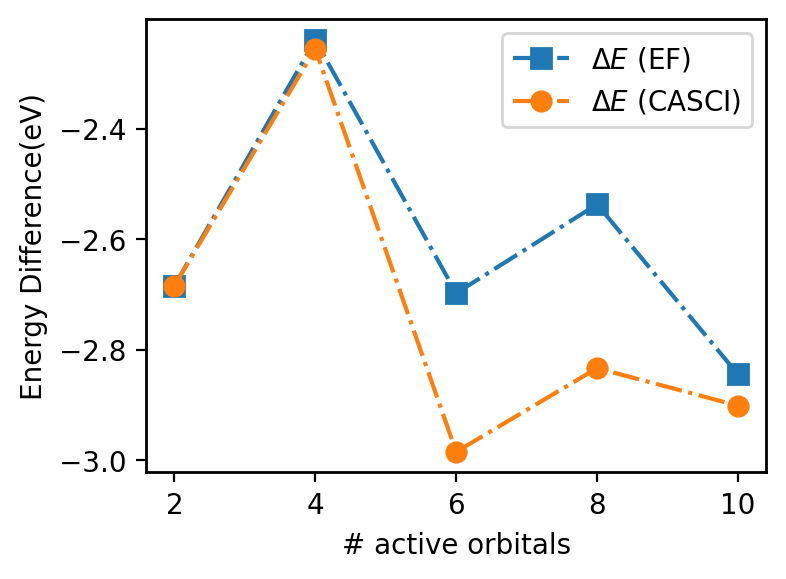}
	\caption{Energy difference between the product and reactant using the EF and CASCI method as a function of the active space size at the $\vett{\Gamma}$-point.}
	\label{fig:EF_ediff}
\end{figure}

\begin{table}[h!]
\centering
\def\arraystretch{1.5}
\begin{tabular}{|c|c|c|c|c|c|}
\hline\hline
\textbf{$\vett{k}$-point} & \textbf{2} & \textbf{4} & \textbf{6} & \textbf{8} & \textbf{10}\\ \hline\hline
$\vett{\Gamma}$ & -3981.02732 & -3981.06264 & -3981.07416 & -3981.10488 & -3981.13650 \\\hline
$\vett{k1}$ & -3980.73290 & -3980.76923 & -3980.80336 & -3980.81200 & -3980.84216\\\hline
$\vett{k2}$ & -3980.73362 & -3980.76991	& -3980.80396 & -3980.81305	& -3980.84271\\\hline
$\vett{k3}$ & -3980.73367 & -3980.76996	& -3980.80296 & -3980.81261	& -3980.84269\\\hline
$\vett{k4}$ & -3980.94243 & -3980.97870	& -3981.01488 & -3981.02153	& -3981.05147\\\hline
$\vett{k5}$ & -3980.94243 & -3980.97870	& -3981.01488 & -3981.02153	& -3981.05147\\\hline
$\vett{k6}$ & -3980.94173 & -3980.97796	& -3981.01424 & -3981.02079	& -3981.05057\\\hline
$\vett{k7}$ & -3980.85478 & -3980.89114	& -3980.92797 & -3980.93560	& -3980.96345\\\hline
$\vett{k8}$ & -3980.85469 & -3980.89103	& -3980.92780 & -3980.93529	& -3980.96344\\\hline
$\vett{k9}$ & -3980.94167 & -3980.97795	& -3981.01411 & -3981.02048	& -3981.05070\\\hline
$\vett{k10}$ & -3980.85432 & -3980.89067 & -3980.92764 & -3980.93656 & -3980.96303\\\hline
$\vett{k11}$ & -3980.85431 & -3980.89066 & -3980.92762 & -3980.93653 & -3980.96302\\\hline
$\vett{k12}$ & -3980.94173 & -3980.97796 & -3981.01424 & -3981.02079 & -3981.05057\\\hline
$\vett{k13}$ & -3980.94167 & -3980.97795 & -3981.01411 & -3981.02048 & -3981.05070\\\hline
$\vett{k14}$ & -3980.85467 & -3980.89100 & -3980.92789 & -3980.93613 & -3980.96339\\\hline
$\vett{k15}$ & -3980.85471 & -3980.89105 & -3980.92786 & -3980.93493 & -3980.96343 \\\hline \hline
\end{tabular}
\caption{CASCI energies in Hartree for the reactant system for active space sizes of 2 to 10 natural orbitals (along columns). Each row corresponds to a $\vett{k}$-point of the $4\times4\times1$ Monkhorst-Pack grid. They are arranged such that top $y^2$ rows correspond to $y\times y\times1$ $\vett{k}$-points for $y \in \{1,2,4\}$}
\label{CASCI_reactant}
\end{table}

\begin{table}[h!]
\centering
\def\arraystretch{1.5}
\begin{tabular}{|c|c|c|c|c|c|}
\hline\hline
\textbf{$\vett{k}$-point} & \textbf{2} & \textbf{4} & \textbf{6} & \textbf{8} & \textbf{10}\\ \hline\hline
$\vett{\Gamma}$ & -3981.12598 & -3981.14553 & -3981.18386 & -3981.20901 & -3981.24312 \\\hline
$\vett{k1}$ & -3980.84116 & -3980.86099 & -3980.90073 & -3980.92480 & -3980.95919\\\hline
$\vett{k2}$ & -3980.84021 & -3980.86008 & -3980.89705 & -3980.92330 & -3980.95759\\\hline
$\vett{k3}$ & -3980.84122 & -3980.86071	& -3980.89915 & -3980.92476 & -3980.95931\\\hline
$\vett{k4}$ & -3981.01493 & -3981.03404	& -3981.07256 & -3981.09781 & -3981.13191\\\hline
$\vett{k5}$ & -3981.01493 & -3981.03404	& -3981.07256 & -3981.09781 & -3981.13191\\\hline
$\vett{k6}$ & -3981.02091 & -3981.03972	& -3981.07846 & -3981.10327 & -3981.13760\\\hline
$\vett{k7}$ & -3980.94182 & -3980.96125	& -3980.99898 & -3981.02423 & -3981.05856\\\hline
$\vett{k8}$ & -3980.93851 & -3980.95933	& -3980.99836 & -3981.02129 & -3981.05546\\\hline
$\vett{k9}$ & -3981.01783 & -3981.03635	& -3981.07467 & -3981.10059 & -3981.13482\\\hline
$\vett{k10}$ & -3980.94069 & -3980.96021 & -3980.99808 & -3981.02322 & -3981.05760\\\hline
$\vett{k11}$ & -3980.94069 & -3980.96021 & -3980.99808 & -3981.02322 & -3981.05760\\\hline
$\vett{k12}$ & -3981.02091 & -3981.03972 & -3981.07846 & -3981.10327 & -3981.13760\\\hline
$\vett{k13}$ & -3981.01783 & -3981.03634 & -3981.07466 & -3981.10059 & -3981.13482\\\hline
$\vett{k14}$ & -3980.93851 & -3980.95933 & -3980.99835 & -3981.02129 & -3981.05546\\\hline
$\vett{k15}$ & -3980.94182 & -3980.96125 & -3980.99898 & -3981.02423 & -3981.05856\\\hline \hline
\end{tabular}
\caption{CASCI energies in Hartree for the product system for active space sizes of 2 to 10 (along columns). Each row corresponds to a $\vett{k}$-point of the $4\times4\times1$ Monkhorst-Pack grid. They are arranged such that top $y^2$ rows correspond to $y\times y\times1$ $\vett{k}$-points for $y \in \{1,2,4\}$}
\label{CASCI_product}
\end{table}
\clearpage

\subsection*{Details of Circuit Reduction}
\label{sec:S6}

\TG{
In this section, we present additional details about the circuit reduction technique. We outline the steps of the procedure and provide an illustration for a 6-qubit circuit.

\paragraph{Nomenclature.}
Following the main text, we consider QCC quantum circuits with $n$ qubits prepared in an initial bitstring state $\psi_{\mathrm{HF}}$. These circuits are subject to exponentials of Pauli operators $P_1 \dots P_m$ and the measurement of a Hermitian operator $O$.

\paragraph*{Initial qubit permutation.}
As a preliminary step to the circuit reduction technique, we apply a qubit permutation to prioritize the action of Pauli operators $P_k$ on the rightmost qubits whenever possible, giving precedence to $P_k$ over $P_{k+1}$. Starting with $P_1$, we permute the qubits so that $P_1 = \mathbf{1}_{n-w_1} \otimes R_1$, where $P_1$ acts non-trivially on $w_1$ out of the $n$ qubits, which are the rightmost in the register. This permutation is also applied to the initial state, other Pauli operators in the ansatz, and the observable of interest. The same procedure is repeated for the subsequent Pauli operators, such as $P_2 = L_2 \otimes R_2$, where we separately permute the first $n-w_1$ qubits and the last $w_1$ qubits to ensure $L_2$ acts non-trivially on the rightmost qubits. This process is applied to all $m$ Pauli operators in the ansatz.

We note that:
(i) Finding and applying this qubit permutation incurs a polynomial cost in the number of qubits, not factorial.
(ii) Empirical observations indicate that permuting qubits leads to more efficient quantum circuits after reduction. However, users may choose to compare the performance of circuit reduction with and without initial qubit permutation on a case-by-case basis.

\paragraph*{Circuit simplification lemmas.}
The circuit reduction technique utilizes the identities shown in Supplementary Figure \ref{fig:circuit_identities} to decompose the circuit of an exponential of a given Pauli operator into a Clifford-only part and a part that includes non-Clifford gates, denoted as $V_k(\theta_k) = C_k U_k(\theta_k)$. It is important to note that the Clifford-only part $C_k$ follows the non-Clifford part $U_k(\theta_k)$.

\begin{figure}[h!]
	\centering
	\includegraphics[width=\textwidth]{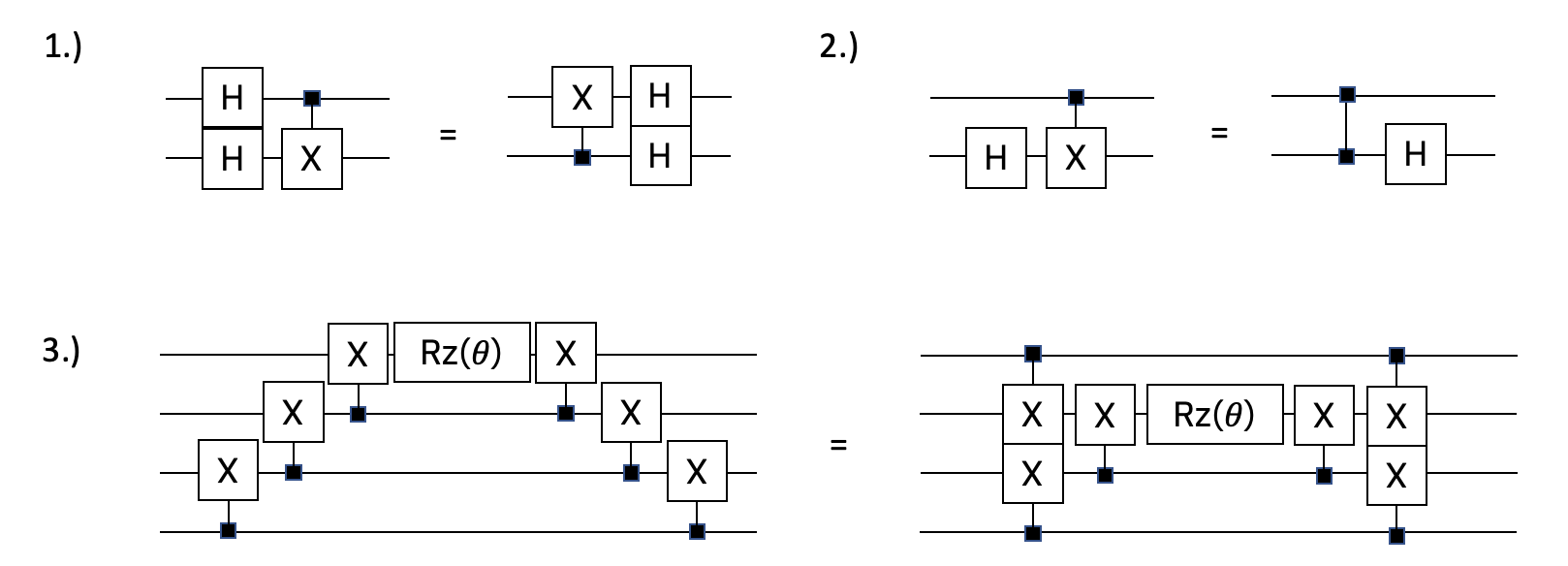}
	\caption{Circuit identities used in the circuit reduction protocol.}
	\label{fig:circuit_identities}
\end{figure}

\paragraph*{Removal of Clifford transformations.}

The exponential of a Pauli operator, $V_k(\theta_k) = C_k U_k(\theta_k)$, can be simplified by removing the Clifford part. This is accomplished by applying the Clifford transformation, $C_k$, to all Pauli operators that follow $P_k$ in the ansatz and to the observable of interest. Specifically, $P_l$ is transformed to $C_k^\dagger P_l C_k$ for $l \geq k$, and $O$ is transformed to $C^\dagger O C$. It is worth noting that this redefinition can be performed efficiently on a classical computer since $C_1$ is a Clifford circuit, $P_k$ is a Pauli operator, and $O$ is a linear combination of Pauli operators.

\paragraph*{A complete example.}

As an example, we consider the application of the circuit reduction method to a QCC ansatz with initial state $|\psi_{\mathrm{HF}} \rangle = |100100\rangle$ and Pauli strings $P_{1}=\mathrm{YXXXXX}$, $P_{2}= \mathrm{IYIIXI}$, and $P_{3}=\mathrm{IYIXXX}$. In the preliminary step, qubits are permuted as per paragraph $(a)$. After permutation, the Pauli operators and the initial bitstring state become $P_{1}=\mathrm{XYXXXX}$, $P_{2}= \mathrm{IIIIXY}$, $P_{3}=\mathrm{IIXXXY}$, and $|\psi_{\mathrm{HF}} \rangle = |010100\rangle$, respectively. The expectation value of the observable $O$ is given by
\begin{eqnarray}
\langle O\rangle &= &\langle \psi_{\mathrm{HF}}|e^{i\theta_{1}P_{1}}e^{i\theta_{2}P_{2}}e^{i\theta_{3}P_{3}}|O|e^{-i\theta_{3}P_{3}}e^{-i\theta_{2}P_{2}}e^{-i\theta_{1}P_{1}}|\psi_{\mathrm{HF}}\rangle\\
&=& \langle 0|V_{0}^{\dagger}V_{1}^{\dagger}(\theta_{1})V_{2}^{\dagger}(\theta_2)V_{3}^{\dagger}(\theta_3)|O|V_{3}(\theta_3)V_{2}(\theta_2)V_{1}(\theta_1)V_{0}|0 \rangle\;,
\end{eqnarray}
where
\begin{eqnarray}
|\psi_{\mathrm{HF}}\rangle &=&V_{0}|0\rangle\\
V_{k}(\theta_{k})&=& e^{-i\theta_{k}P_{k}} ~~\text{for $k = 1,2,3$}
\end{eqnarray}
}
\TG{
We iteratively decompose each quantum circuit $V_k(\theta_k)$ into a non-Clifford circuit $U_k(\theta_k)$ followed by a purely Clifford circuit $C_k$, as shown in Supplementary Figure \ref{fig:circuit_reduction}. For the first Pauli, the circuit $e^{-i\theta_{1}P_{1}}|\psi_{\mathrm{HF}}\rangle = V_{1}(\theta_{1})V_{0}|0\rangle$ separates into $V_{1}(\theta_{1})V_{0}|0\rangle = C_{1}U_{1}(\theta_{1})|0\rangle$, with $C_{1}$ represented by the solid orange box in Supplementary Figure \ref{fig:circuit_reduction} (panel 1) and $U_{1}(\theta_{1})$ represented by the dashed green box. The Clifford circuit $C_{1}$ transforms the remaining Pauli operators and the Hermitian operator $O$ as follows,
\begin{eqnarray}
\langle O\rangle &=&
\langle 0|U_{1}^{\dagger}(\theta_{1})C^{\dagger}_{1} V_{2}^{\dagger}(\theta_2) V_{3}^{\dagger}(\theta_3) |O| V_{3}(\theta_3) V_{2}(\theta_2) C_{1} U_{1}(\theta_{1})|0 \rangle\\
&=& \langle 0|U_{1}^{\dagger}(\theta_{1}) C^{\dagger}_{1} V_{2}^{\dagger}(\theta_2) C_{1} C_{1}^{\dagger} V_{3}^{\dagger}(\theta_3) C_{1} |C_{1}^{\dagger} O C_{1} | C_{1}^{\dagger} V_{3}(\theta_3) C_{1} C_{1}^{\dagger} V_{2}(\theta_2) C_{1} U_{1}(\theta_{1})|0 \rangle\\
&=& \langle 0|U_{1}^{\dagger}(\theta_{1}) V_{2}^{'\dagger}(\theta_2) V_{3}^{'\dagger}(\theta_3) |O'| V'_{3}(\theta_3) V'_{2}(\theta_2) U_{1}(\theta_{1})|0 \rangle
;,
\end{eqnarray}
where
\begin{eqnarray}
V'_{k}(\theta_{k}) &=& C_{1}^{\dagger} V_{k}(\theta_{k}) C_{1} = e^{-i\theta_{k}C_{1}^{\dagger}P_{k}C_{1}} = e^{-i\theta_{k}P'_{k}}~~\text{for $k = 2,3$} \\
O' &=& C_{1}^{\dagger} O C_{1} ;.
\end{eqnarray}
We then consider the circuit $V'_{2}(\theta_{2})$.
This circuit is decomposed into a Clifford part and non-Clifford part as shown in Supplementary Figure~\ref{fig:circuit_reduction} (panel 3), i.e. $V'_{2}(\theta_{2}) = C_2 U_{2}(\theta_{2})$. The Clifford transformation $C_2$ is applied to the third Pauli operator and the Hermitian operator $O^\prime$ as follows,
\begin{eqnarray}
V''_{k}(\theta_{k}) &=& C_{2}^{\dagger}V'_{k}(\theta_{k})C_{2} = e^{-i\theta_{k}P''_{k}}~~\text{for $k = 3$}\\
O'' &=& C_{2}^{\dagger}O'C_{2} \;.
\end{eqnarray}
The quantum circuit implementing the exponential of the third Pauli operator, $V''_{3}(\theta_{3})$, is decomposed into a Clifford and non-Clifford part, $V_{3}''(\theta_{3}) = C_{3}U_{3}(\theta_{3})$. The process terminates with the application of the final Clifford transform obtained in Supplementary Figure~\ref{fig:circuit_reduction} (panel 5) to the measurement operator, i.e. $O''' = C_{3}^{\dagger}O''C_{3}$. Thus, the quantum circuit after the circuit reduction process has substantially lower depth and fewer CNOT gates and qubits, as shown in Supplementary Figure~\ref{fig:circuit_reduction} (panel 6).} 

\begin{figure}[h!]
	\centering
	\includegraphics[width=0.68\textwidth]{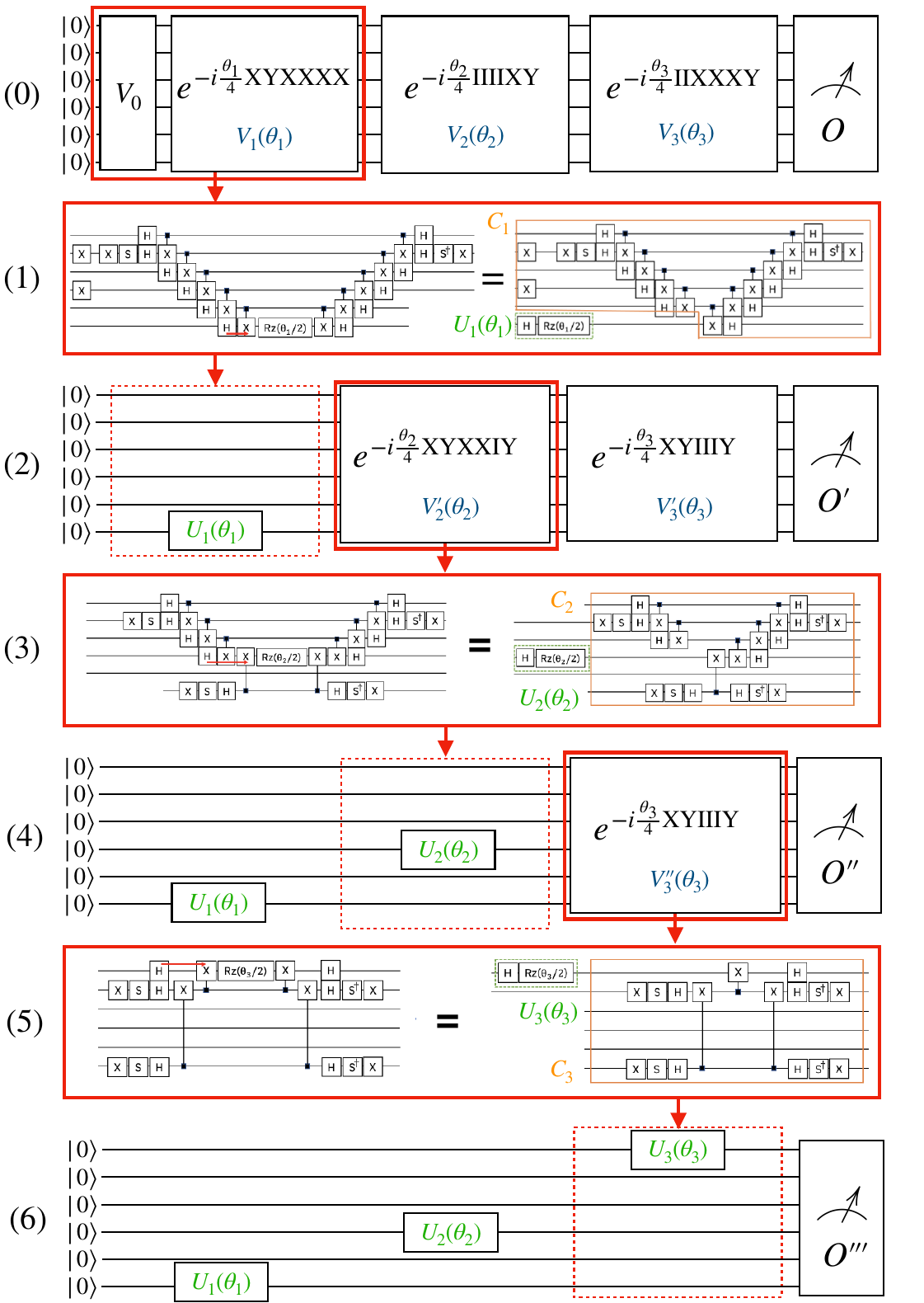}
	\caption{\TG{Circuit reduction applied step by step. Step 0: initial circuit, comprising initialization in the bitstring $| 000000 \rangle$ with unitary $V_{0}$, application of the exponentials of three Pauli operators $P_{1}=\mathrm{XYXXXX}$, $P_{2}=\mathrm{IIIIXY}$, $P_{3}=\mathrm{IIXXXY}$, and measurement of a Hermitian operator $O$.
Step 1: simplification (left) of $\exp(-i \theta_1/4 P_1)$ based on known circuit identities and partition (right) of the simplified circuit into a reduced non-Clifford circuit $U_1(\theta_1)$ (green dashed box) and a Clifford circuit $C_1$ (orange solid box)
Step 2: the Clifford circuit $C_1$ in Step 1 is applied to the Pauli operators $P_2$ and $P_3$ and the Hermitian operator $O$.
Points 3 is the same as point 1 for the unitary transformation $\exp(-i \theta_2/4 P_2^\prime)$.
Point 4 is the same as point 2 for the unitary transformation with $C_{2}$.
Point 5 is the same as point 1 for the unitary transformation $\exp(-i \theta_3/4 P_3^{\prime\prime})$.
Point 6 shows the final circuit, consisting of the measurement of a transformed operator $O^{\prime\prime\prime}$ on the state $U_3(\theta_3) U_2(\theta_2) U_1(\theta_1) | 000000 \rangle$.}}
    \label{fig:circuit_reduction}
\end{figure}

\clearpage
\subsection*{Details of Hardware Calculations}
Circuit reduction techniques employing exact circuit identities and Clifford transformations were used to reduce the required circuit resources for the QCC ansatzes with 2 and 5 Pauli strings. The circuits which were run on the quantum hardware for 2 and 10 natural orbitals active spaces are given in Supplementary Figure \ref{fig:QCC_circ_2NO} and \ref{fig:QCC_circ_10NO} respectively.

\TG{Note that, without loss of generality, we have used a slightly different Clifford transformation to diagonalize the $Y$ Pauli in Supplementary Figure \ref{fig:circuit_reduction} and Supplementary Figures (\ref{fig:QCC_circ_2NO}-\ref{fig:QCC_circ_10NO}). While the typical circuit identity to diagonalize the $Y$ Pauli operator is $Y = C^{\dagger}ZC$, where $C=HSX$ as shown in Supplementary Figure \ref{fig:circuit_reduction}, we used the alternative identity $Y = C^{\dagger}(-Z)C$ where $C = HS$ and absorbed the negative sign during the parameter optimization in VQE.}

\begin{figure}[h!]
	\centering
	\includegraphics[width=0.9\textwidth]{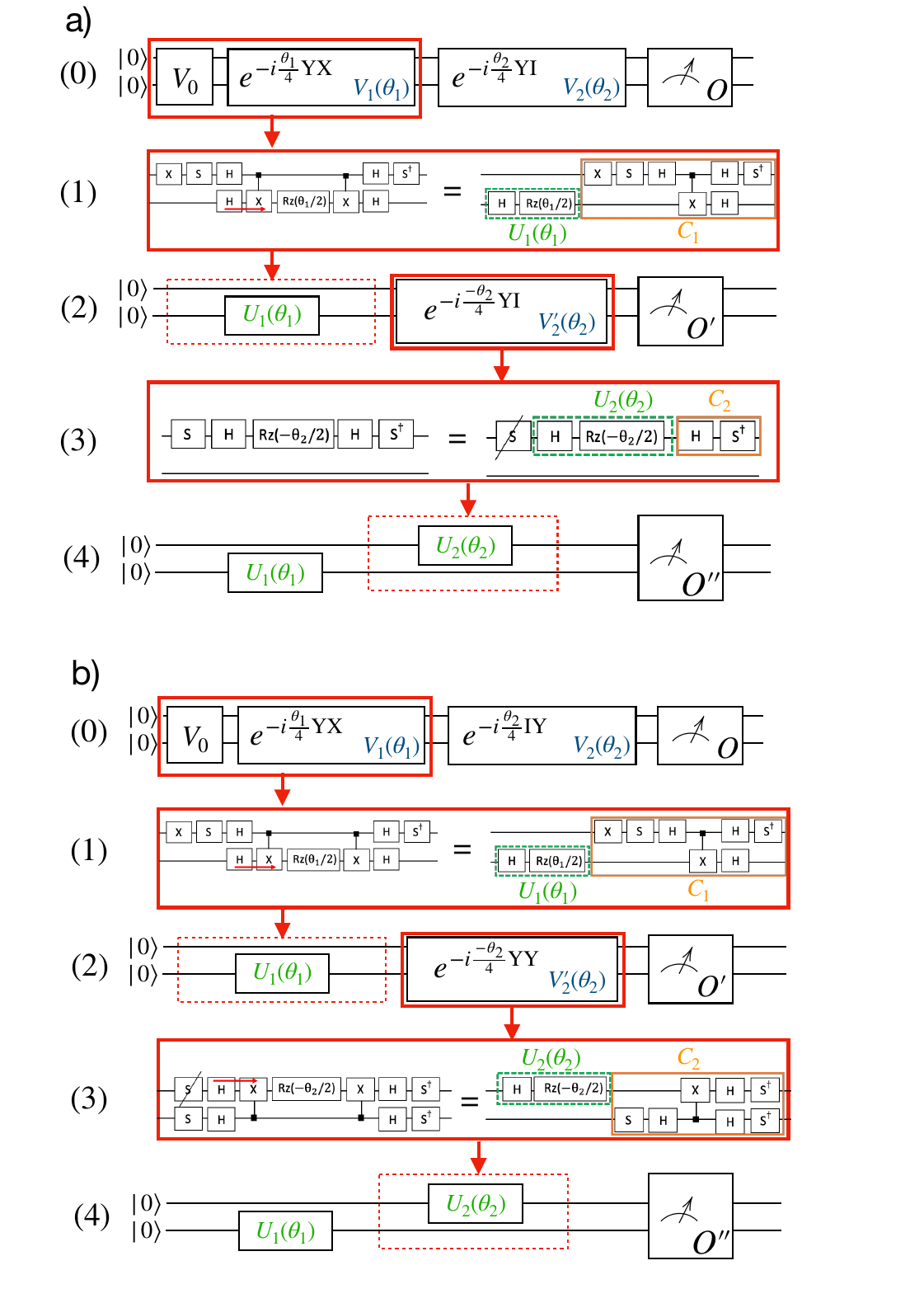}
	\caption{\TG{Detailed steps of the circuit reduction} for QCC ansatzes pertaining to the reactant (a) and product (b). The ansatzes use 2 Pauli strings in a two-orbital active space. \TG{The identities from Supplementary Figure (\ref{fig:circuit_identities}) and the fact that $S|0\rangle = |0\rangle$ were used in determination of the Clifford circuits. In this 2-qubit case, it was not necessary to permute qubits before the circuit reduction.}}
	\label{fig:QCC_circ_2NO}
\end{figure}

 \begin{figure}[h!]
	\centering
	\includegraphics[width=\textwidth]{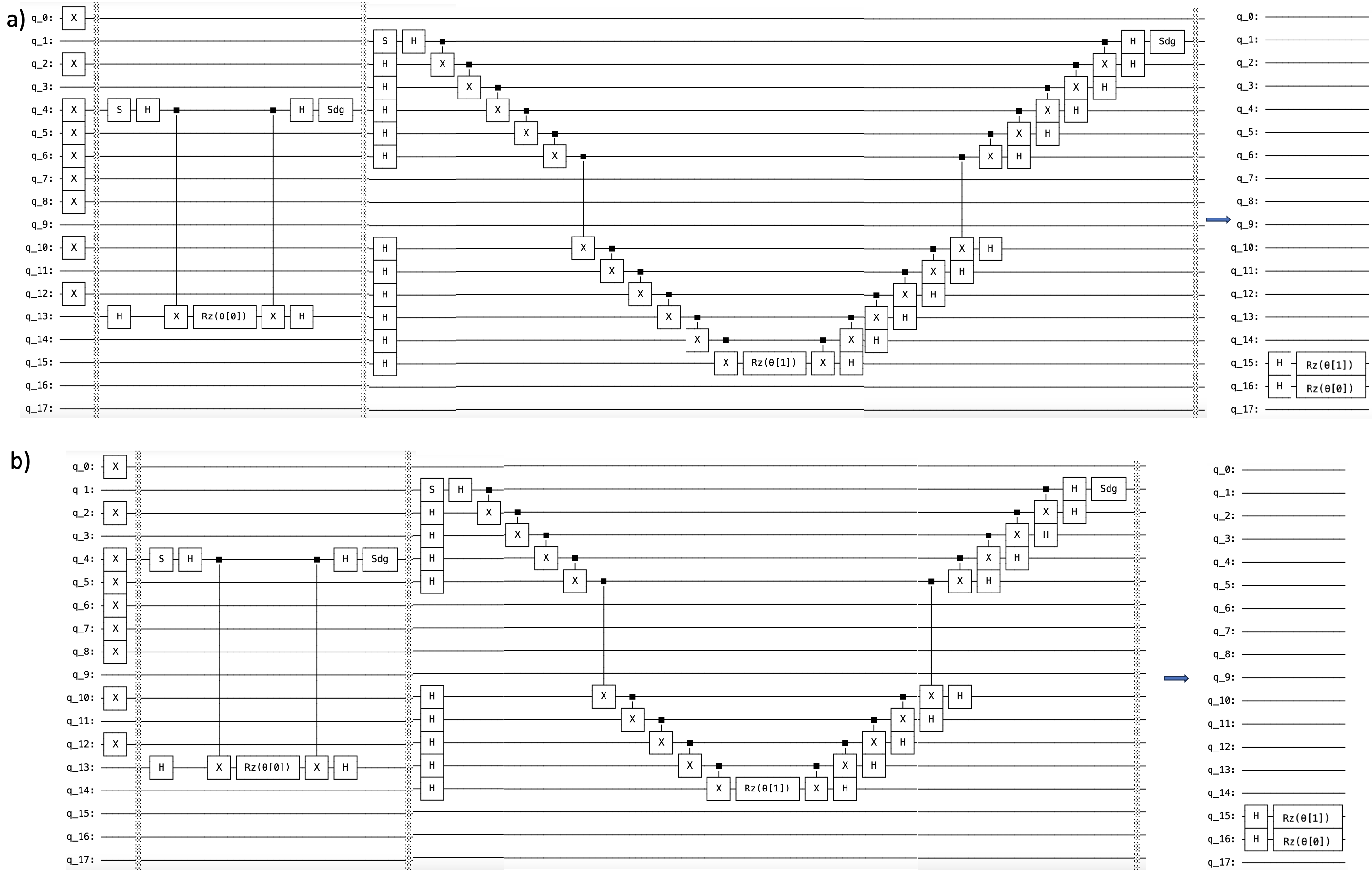}
	\caption{Original and reduced circuits implementing QCC ansatzes for the reactant (a) and product (b). The ansatzes use 2 Pauli strings in a 10-orbital active space. \TG{In (a) the Paulis and initial bit-string state before permutation are $P_{1} = \mathrm{IIIIYIIIIIIIIXIIII}$, $P_{2} = \mathrm{IYXXXXXIIIXXXXXXII}$ and $|\psi_{\mathrm{HF}}\rangle = |101011111010100000\rangle$. In (b) Paulis and initial bit-string state before qubit permutation are $P_{1} = \mathrm{IIIIYIIIIIIIIXIIII}$, $P_{2} = \mathrm{IYXXXXIIIIXXXXXIII}$ and $|\psi_{\mathrm{HF}}\rangle = |101011111010100000\rangle$. The quantum circuits before qubit permutation and circuit reduction are shown on the left-hand side of panels (a) and (b). After applying the circuit reduction technique as illustrated in example Supplementary Figure (\ref{fig:circuit_reduction} and \ref{fig:QCC_circ_2NO}), the reduced quantum circuits are shown on the right-hand side in the panels (a) and (b).}}
	\label{fig:QCC_circ_10NO}
\end{figure}

\clearpage

\section{Supplementary References}